\documentclass[preprint,showkeys]{revtex4}
\usepackage{amssymb,amsmath}
\usepackage{epsfig}
\font\bba=msbm10 
\font\bbb=msbm8 
\font\bbc=msbm6 
\newfam\bbfam
\textfont\bbfam=\bba
\scriptfont\bbfam=\bbb
\scriptscriptfont\bbfam=\bbc
\def\bb{\fam\bbfam\bba}
\def\R{{\bb R}}

\DeclareMathOperator{\Tr}{Tr}
\newcommand{\tarc}{\mbox{\large$\frown$}}
\newcommand{\arc}[1]{\stackrel{\tarc}{#1}}
\begin{document}
\title{Electrostatics on the sphere  with applications
to Monte Carlo simulations of two dimensional  polar fluids.}
\author{Jean-Michel Caillol}
\email{Jean-Michel.Caillol@th.u-psud.fr}
\affiliation{University of  Paris-Sud, CNRS, LPT, UMR 8627, Orsay, F-91405, France}                                             %
\date{\today}        
\begin{abstract}
We present  two methods for solving the electrostatics of point charges and multipoles
on the surface of a sphere, \textit{i.e.} in the space $\mathcal{S}_{2}$, with applications to numerical simulations of
two-dimensional polar fluids. 

In the first approach, point charges are associated
with uniform neutralizing backgrounds to form neutral pseudo-charges, while, in the second, one instead
considers bi-charges, \textit{i.e.} dumbells of antipodal point charges of  opposite signs. 
We establish the expressions of the electric potentials  of pseudo-  and bi-charges as isotropic solutions of 
the Laplace-Beltrami equation in $\mathcal{S}_{2}$. A multipolar expansion of pseudo- and bi-charge 
potentials leads to  the electric potentials of mono- and bi-multipoles respectively. 
These potentials  constitute
non-isotropic solutions of the Laplace-Beltrami equation the general solution of which in spherical coordinates 
is recast under a new appealing form.

We then focus on  the case of  mono- and bi-dipoles and build the theory of dielectric media in $\mathcal{S}_{2}$.
We notably obtain the expression of the static dielectric constant of a uniform isotropic polar fluid living
in $\mathcal{S}_{2}$ in term  of the polarization fluctuations of subdomains of  $\mathcal{S}_{2}$.
We also derive the long range behavior of the equilibrium pair correlation function 
under the assumption that it is  governed  by macroscopic electrostatics.
These theoretical developments  find their application in  Monte Carlo simulations of  the $2D$  fluid of
dipolar hard spheres.
 Some preliminary numerical experiments  are
discussed with a special emphasis on finite size effects, a careful  study of the thermodynamic limit, and a check
of the theoretical predictions for the asymptotic behavior of the  pair correlation function.
\end{abstract}
\keywords{Laplace-Beltrami equation on the sphere; Two-dimensional  polar fluids; Monte Carlo simulations.}
\maketitle                                                                                
\section{Introduction}
\label{Introduction}
The idea of using the two dimensional ($2D$) surface of a  sphere, \textit{i.e.} the space 
$\mathcal{S}_{2}$,  to perform numerical
simulations of a  $2D$ fluid phase can be tracked back to a paper 
by J.-P. Hansen {\em et al.} devoted to a study of the electron gas at 
the surface of liquid Helium \cite{JPH}. 
Subsequently, the same idea was used to study the crystallization of the $2D$
one-component plasma (with $\propto \log r$ interactions) \cite{2DOCP}, to establish the phase
diagram of the $2D$ Coulomb gas \cite{2DTCP},  and to determine
some thermodynamic and structural properties of the $2D$ polar fluid (with $\propto 1/r^2$ interactions) in its liquid
phase \cite{these,Orsay_1}. The generalizations to $3D$ systems, implying the use of the surface of a
 $4D$ hypersphere, \textit{i.e.} the space  $\mathcal{S}_3$, is due to Caillol and Levesque \cite{Orsay_2}. Some 
improvements on this early work, mostly applications concerning   $3D$ polar fluids,
were  published recently \cite{Tr_Cai}.

Several simple ideas can be put forward to justify the use of a hyperspherical geometry in numerical simulations of plasmas and
Coulombic fluids (\textit{i.e.} fluids made of ions and/or dipoles) :
\begin{itemize}
 \item{(i)} 
The $n$-dimensional non-Euclidian space $\mathcal{S}_n$ (in practice $n=2,3$), albeit finite, is homogeneous 
and isotropic, in the sense that it is invariant under the group $\mathcal{O}(n+1)$ of the $(n+1) D$
rotations of the Euclidian space $E_{n+1}$; it is thus well suited for the simulation of a fluid phase (in the bulk). 
 \item {(ii)} 
The laws of electrostatics
can easily be obtained in $\mathcal{S}_n$ and the Green's function of
Laplace-Beltrami equation (\textit{i.e.} Coulomb potential) is known~\cite{Cohl}. It  has
a very simple analytical expression, tailor-made  for numerical evaluations.
 \item {(iii)} The inclusion of charged or uncharged walls can easily be done in order to study
the structure of liquids, plasmas, or collo\"{\i}ds at interfaces \cite{Orsay_2,Pellenq_1,Pellenq_2,Pellenq_3,Mazars} 
\end{itemize}

The second point (ii) was partly overlooked by the authors of Ref.~\cite{these,Orsay_1} who used an empirical
$2D$ dipole/dipole interaction, unfortunately \textit{not} deduced from a solution of Laplace-Beltrami equation
in  $\mathcal{S}_{2}$,
which casts some doubts on the validity of their  results  concerning  the $2D$ Stockmayer fluid,
notably those concerning its dielectric properties.

Some years after we are now in position to propose in this paper  two possible, both rigorously correct,
dipole/dipole pair-potentials and to use both 
of them into actual numerical simulations of a  $2D$ dipolar hard sphere (DHS) fluid.
Here, not only we  extend to  $\mathcal{S}_{2}$  the   recent  developments on $3D$
polar fluids in $\mathcal{S}_3$ discussed in Ref.~\cite{Tr_Cai}, but we also
present additional new results on multipolar  expansions in $\mathcal{S}_{2}$, 
not yet available in $\mathcal{S}_{3}$, in Sec.~\ref{Electrostatics}.

This article is thus a contribution to the  physics of $2D$ fluids, collo\"{\i}ds  or plasmas, the atoms or molecules of
which interact \textit{via} electrostatic  pair potentials derived  from
a solution of the $2D$ Laplace equation.
We emphasize that the systems that we  consider   cannot be
seen as thin layers of  real  $3D$ systems. In this  case, the electrostatic interactions 
should be derived from  the solutions of the $3D$ Laplace equation.
Simulations of such $3D$ systems  constrained to live in a $2D$ geometry
are nevertheless also possible in spherical geometries. In that case one should consider
a $3D$ fluid of  $\mathcal{S}_3$ made of particles interacting by pair potentials deduced from 
solutions of Laplace-Beltrami equation in  $\mathcal{S}_3$ and constrained to stay on the equator of the hypersphere.
This geometrical locus indeed
identifies with 
the sphere $\mathcal{S}_2$. This approach was used for  the  $3D$ restricted primitive model of electrolytes
in Ref.~\cite{Orsay_3}.

In this work our aim is to demonstate the validity of the sphere $\mathcal{S}_{2}$
to perform  Monte Carlo (MC) simulations of a  $2D$ polar fluid.
We note that, although  $2D$ dipolar
fluids do not exist \textit{per se} in nature, the model  could be used via various mappings for
applications as, recently, for the hydrodynamics of two-dimensional
microfluids of droplets \cite{Nature}. 

Our paper is organized as follows.  After this introduction (Sec.~\ref{Introduction}) 
we discuss in depth the electrostatics of the distributions of charges in  $\mathcal{S}_{2}$
and their multipole expansions.
The material reported in  Sec.~\ref{Electrostatics}  is an intricate mixture of old and new stuff. 
The underlying idea is the  remark of Landau and  Lifchitz in  Ref.~\cite{Landau}  that, in a finite space 
such as  $\mathcal{S}_{2}$,
Laplace-Beltrami equation admits  solutions if and only if  the total
electric charge of the space is equal to zero. Therefore, in  $\mathcal{S}_{2}$, the building brick of electrostatics cannot
be a single point charge $q$ as  in the ordinary $2D$ Euclidian space $E_2$.
We are led instead to consider  a pseudo-charge \cite{Orsay_2}, a neologism denoting the 
association of a point charge and a uniform neutralizing background of opposite charge.
Alternatively  we can consider  a bi-charge, \textit{i.e.} 
a neutral dumbell made  of  two antipodal charges of opposite signs $+q$ and $-q$ as in Ref.~\cite{Caillol_bi}. 
Both approaches yield  independent electrostatics theories  which are sketched
in Sec.~\ref{Electrostatics} and~\ref{Dipoles}.

The case of dipoles is then examined more specifically in Sec.~\ref{Dipoles}. We need make a distinction between mono-dipoles,
which are obtained as the leading term of a  multipole expansion of a neutral set of
pseudo-charges, and bi-dipoles, obtained
in a similar way  from bi-charges. As a consequence
two possible microscopic models of dielectric media can be worked out, those made of mono-dipoles
living on the entire sphere $\mathcal{S}_2$ and those  made of bi-dipoles
living on the northern hemisphere $\mathcal{S}_2^+$.

These two types of models  are then studied in the framework of Fulton's theory~\cite{Fulton_I,Fulton_II,Fulton_III,Caillol_2_a}
in  Sec.~\ref{2D-Polar-Fluid}. Fulton's theory realizes a harmonious synthesis of linear response theory and the 
macroscopic theory of dielectric media.
It allows us to obtain 
\begin{itemize}
 \item   a family of formula for the dielectric constant $\epsilon $ of the fluid, well adapted for its
determination in MC simulations.
\item an expression for the asymptotic pair correlation in  $\mathcal{S}_{2}$ under the assumption that it should
be dictated by the laws of macroscopic electrostatics.
\end{itemize}

Monte-Carlo (MC) simulations of the $2D$ DHS fluid in the isotropic fluid phase
are then discussed in Sec.~\ref{Simulations}. We have chosen to report MC
data only  for  two thermodynamic states, both
in  the isotropic fluid phase and extensive MC simulations of the  $2D$ DHS fluid will be published elsewhere.
These states could serve
as benchmarks for future numeric studies. One of these states was studied many years ago in Ref.~\cite{Morris} by means of MC
simulations  within  standard periodic boundary conditions (see also \cite{Perram}). Comparisons of our results with
those obtained in  this pioneer work are OK. 
A carefull finite size scaling study of our data yields results of a high 
precision, notably for  the energy, probably unattainable by standard simulation methods.

Conclusions are drawn in  Sec~\eqref{Conclu}.

\section{Electrostatics of $2D$ Charges and Multipoles}
\label{Electrostatics}
\subsection{The Plane}
\label{plane}
Let us first recall 
that the electrostatic potential $V_{E_2}(\boldsymbol{\rho})$ at
a source-free point $\boldsymbol{\rho}=(x,y)$ of the Euclidian plane $E_2$ satisfies the 2D Laplace
equation $\Delta_{\mathrm{E}_2} V_{E_2} =0$, which in polar coordinates $(\rho,\varphi)$ reads
\begin{equation}
\label{Lap1}
\Delta_{\mathrm{E}_2} V_{E_2} =  \frac{1}{\rho}  \frac{\partial}{\partial \rho}  \left( \rho   \frac{\partial V }{\partial \rho}          \right)  + 
 \frac{1}{ \rho^2} \frac{\partial^{\,2} V}{\partial \varphi^2} =0 \; .
\end{equation}
The general solution of Eq.~\eqref{Lap1} is \cite{Jackson,Joslin}
\begin{equation}
 \label{Sol_E2}
V_{E_2}(\rho, \varphi) = a_0 + b_0 \log \rho + \sum_{n=1}^{\infty} a_n \rho^n \cos(n \varphi + \alpha_n)
+ \sum_{n=1}^{\infty} b_n \rho^{-n} \cos(n \varphi + \beta_n) \; ,
\end{equation}
where $a_n$, $b_n$, $\alpha_n$, and $\beta_n$ are arbitrary constants. 
The terms of the \textit{r.h.s.} of~\eqref{Sol_E2} which are singular at $\rho=0$ may be interpreted
as the potentials created by point multipoles located at the origin $O$, while those singular at  $\rho=\infty$
as  the potentials of point multipoles at infinity. In $E_2$ the potential of a point charge
$q$ located at $O$ is $-q \log \rho$ up to an additional constant; 
it is the Green's function of Eq.~\eqref{Lap1} in  $E_2$
(without boundaries) and it satisfies the 2D Poisson's equation
\begin{equation}
  \label{Poisson}
\Delta_{\mathrm{E}_2} \left( -\log \rho \right) = -2 \pi \delta_{\mathrm{E}_2}(M,M_0) \equiv - 2 \pi \delta(x) \delta(y) \; ,
\end{equation}
where $\delta(x) $ denotes the 1D Dirac's  distribution.

Finally, the Green's function 
$-\log (\arrowvert \boldsymbol{\rho} - \boldsymbol{\rho}^{'}\arrowvert )$ can be expanded in polar coordinates
 as \cite{Jackson}
\begin{equation}
\label{Expand1}
-\log (\arrowvert \boldsymbol{\rho} - \boldsymbol{\rho}^{'}\arrowvert ) =
-\log \rho_{<}+ \sum_{n=1}^{\infty}\frac{1}{n} \left( \frac{\rho_<}{\rho_>} \right)^n \cos \left(n \left(\varphi - \varphi^{'}\right) \right) \; ,
\end{equation}
where $\rho_<=\inf(\rho,\rho^{'})$ and  $\rho_>=\sup(\rho,\rho^{'})$.

Eq.~\eqref{Expand1} serves as a starting point for the  $2D$ multipolar expansion of Ref.~\cite{Joslin}.
Let us consider $N$ charges $q_i$ of polar coordinates $(\rho_i,\varphi_i)$ and a point $M$ (coordinates $(\rho,\varphi)$) of
the plane $E_2$ such that $\rho > \rho_i$, $ \forall i $. Making use of~\eqref{Expand1} one finds
\begin{align}
\label{multi1}
 V_{E_2}(M) & = -\sum_{i=1}^N q_i \log (\arrowvert \boldsymbol{\rho} - \boldsymbol{\rho_i}\arrowvert ) \nonumber \\
                    &= V_0  -\frac{1}{2} \sum_{\nu}Q_ {0 \nu} \log \rho +\frac{1}{2}\sum_{n=1}^{\infty} \sum_{\nu} \frac{1}{n \rho^n}Q_ {n \nu} \exp(-i n \nu \varphi) \; ,
\end{align}
where $V_0$ is some unessential constant. In~\eqref{multi1} the index $\nu $ assumes the values $-1$ and $+1$ only. The complex
multipolar moments are defined as
\begin{equation}
 \label{Q1}
Q_ {n \nu} =  \sum_{i=1}^N q_i \rho_i^n \exp(i n \nu \varphi_i)  \; .
\end{equation}
Since $Q_ {n,- \nu}=Q_ {n \nu}^* $ there are \textit{at most}
two independent multipole components at a given order $n$. The potential of a $2D$ multipole of order $n$ therefore
decays as $1/\rho^n$. This remark justifies \textit{a posteriori} the interpretation that we gave
for the terms of the \textit{r.h.s.} of Eq.~\eqref{Sol_E2}.

\subsection{The Sphere}
\label{Sphere}
Here, we examine how the basic laws of $2D$ Euclidian electrostatics swiftly evoked  
in Sec.~\ref{plane}  are changed when the system is wrapped on the surface of a sphere.
Let us first introduce some notations and recall some elementary mathematics.
We denote by  $\mathcal{S}_{2}(O,R)$ the sphere of center $O$ and radius $R$ of  the usual $\mathrm{3D}$ geometry.
It is a (Riemannian) manifold of the $3D$ Euclidean space $E_3$ that we identify with $\R^3$.
When we deal with
the sphere of unit radius we adopt the uncluttered notation $  \mathcal{S}_{2} \equiv   \mathcal{S}_{2}(O,R=1)$.
 Let $M$ be the running point of   $\mathcal{S}_{2}(O,R)$,  we define
the unit vector  $\mathbf{z \in \mathcal{S}_{2}}$ as $\boldsymbol{OM}= R \mathbf{z}$. The 
spherical coordinates  are defined as usual :
$\mathbf{z}=( \sin \theta \cos \varphi, \sin \theta  \sin \varphi, \cos \theta) ^T  $ with
$0 \leq \theta  \leq  \pi$ and $0 \leq \varphi < 2 \pi$. The differential vector 
$d \mathbf{z} = d\theta \,  \mathbf{e}_{\theta}  + \sin \theta d\varphi \, \mathbf{e}_{\varphi} $ 
allows us to obtain two orthogonal unit vectors $(\mathbf{e}_{\theta},\mathbf{e}_{\varphi})$ that span the plane 
$\mathcal{T}_{2}(M)$ tangent to ${\mathcal{S}_{2}}(O,R)$ at point $M$.
Recall that $ \mathbf{e}_{\theta} = (\cos \theta \cos \varphi, \cos \theta \sin \varphi, - \sin \theta)^{T}$ and
 $ \mathbf{e}_{\varphi} = (- \sin \varphi, \cos \varphi, 0)^{T}$. 
In addition and quite specifically since this notion cannot be generalized to higher dimensions,
one has $\mathbf{e}_{\varphi}= \mathbf{z} \times \mathbf{e}_{\theta}$ where 
the symbol $ \times$ denotes the $3D$ vectorial product. Finally the infinitesimal surface element is
$dS = R^2 d\Omega$ where the  infinitesimal solid angle $d \Omega = \sin \theta d \theta  d \varphi$
in spherical coordinates.

In the sequel we will make a repeated use of the unit dyadic tensor 
$\mathbf{U}_{\mathcal{S}_{2}}(\mathbf{z})= \mathbf{e}_{\theta} \mathbf{e}_{\theta} 
+ \mathbf{e}_{\varphi} \mathbf{e}_{\varphi}$  of the plane   $\mathcal{T}_{2}(M)$.
Note that the unit dyadic tensor of Euclidian space $E_3$ is  given by $\mathbf{U}_{E_3}=
\mathbf{U}_{\mathcal{S}_{2}}(\mathbf{z})
+ \mathbf{z} \mathbf{z} $. It is a constant tensor independent of point $M$.
These admittedly old-fashioned objects however allow
an easy definition of the gradient in  $\mathcal{S}_{2}(O,R)$,  or  first differential  Beltrami operator, 
as
\begin{equation*}
 \nabla_{\mathcal{S}_{2}(O,R)}=    \mathbf{U}_{\mathcal{S}_{2}}(\mathbf{z})  \cdot \nabla_{E_3} \; ,
\end{equation*}
where 
$\nabla_{E_3}$ is the usual Euclidian gradient operator of $E_3$ and 
the dot in the r.h.s. denotes the $\mathrm{3D}$ tensorial contraction. 
Note that, obviously,  $\nabla_{\mathcal{S}_{2}  \mathrm{(O,R)}}=  \nabla_{\mathcal{S}_{2}}/R$.

The Laplace-Beltrami operator (or  second differential  Beltrami operator) is defined in a similar way 
as the restriction of the $3D$ Euclidian
Laplacian $\Delta_{E_3}$ to the surface of the sphere \cite{Atkinson}.
\begin{equation}
   \Delta_{\mathcal{S}_{2}(\mathrm{0,R})}  =  \Delta_{E_3} -\frac{\partial^2}{ \partial R^2}
 -\frac{2}{R} \frac{\partial }{\partial R }  \; .
\end{equation}
We have the scaling relation $\Delta_{\mathcal{S}_{2}(\mathrm{0,R})} 
 \equiv  \Delta_{\mathcal{S}_{2}}/ R^2$ and, on the unit sphere, in spherical coordinates 
\begin{equation}
 \Delta_{\mathcal{S}_{2}} = \frac{1}{\sin \theta }  \frac{\partial}{\partial \theta}\left(  \sin \theta   \frac{\partial}{\partial \theta} \right)
     + \frac{1}{\sin^2 \theta} \frac{\partial^2}{\partial \varphi^2}\; .
\end{equation}
 $\Delta_{\mathcal{S}_{2}}$ identifies with minus the squared angular momentum operator of Quantum
Mechanics; therefore it is a Hermitian operator the eigenvectors of which are the $3D$ spherical harmonics
$Y^m_l(\theta,\varphi)$ where $l$ is a non-negative integer and the integer $m$ satisfies $-l \leq m \leq +l$. Moreover
$ \Delta_{\mathcal{S}_{2}} Y^m_l = -l(l+1) Y^m_l $. 

A general solution of Laplace-Beltrami equation in $\mathcal{S}_{2}$, \textit{i.e.}
$ \Delta_{\mathcal{S}_{2}} V_{\mathcal{S}_{2}} =0$, can   
easily be  obtained  in spherical coordinates   by expanding
$V_{\mathcal{S}_{2}}(\theta,\varphi)$ in Fourier series (see \textit{e.g.} \cite{Meteo})
\begin{equation}
 V_{\mathcal{S}_{2}}(\theta,\varphi) = \sum_{m=-\infty}^{m=+\infty} \hat{V}_m(\theta) e^{i m \varphi} \; ,
\end{equation}
which yields for the Fourier coefficients 
\begin{equation}
 \sin \theta \dfrac{d}{d \theta} \sin \theta\dfrac{d}{d \theta}  \hat{V}_m(\theta) = m ^2 \hat{V}_m(\theta) \, .
\end{equation}
In the above equation the change of variables $x = \log  \tan\frac{\theta}{2}$ ($0\leq \theta\leq \pi$) 
leads to the elementary  differential equation
\begin{equation}
  \dfrac{d^2}{d x^2}  \hat{V}_m(x) = m ^2 \hat{V}_m(x)  \; ,
\end{equation}
with the solutions
\begin{subequations}
\begin{eqnarray}
  \hat{V}_0(x) &= a + b x \, , \textrm{ for  }m=0 \; , \\
 \hat{V}_m(x)  &=  K_{\pm m } e^{\pm m x}  \, , \textrm{ for  }m \neq 0 \; ,
\end{eqnarray}
\end{subequations}
where $a$, $b$, and $ K_{\pm m } $ are arbitrary complex constants.
The general real solution of Laplace-Beltrami equation on the sphere can thus be finally written as
\begin{eqnarray}
 \label{Sol_S2}
V_{\mathcal{S}_{2}}(\theta,\varphi) = a_0 + b_0 \log \tan \frac{\theta}{2} + 
     \sum_{n=1}^{\infty} a_n \left[\tan \frac{\theta}{2}\right]^n \cos(n \varphi + \alpha_n) \nonumber \\
+ \sum_{n=1}^{\infty} b_n \left[ \cot \frac{\theta}{2} \right]^n \cos(n \varphi + \beta_n) \; ,
\end{eqnarray}
where $a_n$, $b_n$, $\alpha_n$, and $\beta_n$ are arbitrary real constants. 
The terms of the \textit{r.h.s.} of~\eqref{Sol_S2} which are singular at $\theta=0$ may be interpreted
as the potentials created by point multipoles located at the north pole $\mathcal{N}$ and
 those singular at  $\theta=\pi$
as  the potentials of point multipoles located at the south pole  $\mathcal{S}$.
Remarkably the isotropic term  $ \propto \log ( \tan \theta/2) $ in the \textit{r.h.s.} 
is singular both at $\theta=0$ and $\theta=\pi$ and should identify with the potential of a 
unit bi-charge.

In order to check this assertion, let us first consider  a bi-charge made of a charge $+q$ located
at point $M_0$ of $\mathcal{S}_{2}(\mathrm{0,R})$ and its companion $-q$ located at the antipodal
point $\overline{M}_0$ (with $\overline{\mathbf{z}}_0 = - \mathbf{z}_0$).  The potential $V_{q, M_0}^{\mathrm{bi}}(M)$
at point $M$ is the solution of Poisson's equation
\begin{equation}
\label{Pois_S2}
 \Delta_{\mathcal{S}_{2}} V_{q, M_0}^{\mathrm{bi}}(M) = -2 \pi q  \{   \delta_{\mathcal{S}_{2}}(\mathbf{z}_0, \mathbf{z})
                                                                        -  \delta_{\mathcal{S}_{2}}(\overline{\mathbf{z}}_0, \mathbf{z})   \} \; ,
\end{equation}
where the Dirac's distribution in $\mathcal{S}_{2}$ is defined as 
$ \delta_{\mathcal{S}_{2}}(\mathbf{z}_0, \mathbf{z}) = \delta(1 - \mathbf{z}_0 \cdot \mathbf{z})$ \cite{Atkinson}.
Eq.~\eqref{Pois_S2} is solved by expanding both sides on spherical harmonics; 
after some elementary algebra one finds
\begin{align}
\label{kluk}
 V_{q, M_0}^{\mathrm{bi}}(M) &= q {\sum_{l}\;^{'}} \;  \frac{2 l +1}{l(l+1)} P_l(\cos \psi_{M_0 M}) \nonumber \\
                                                 &= -q -q \log \tan \frac{\psi_{M_0 M}}{2} \; ,
\end{align}
where $P_l(x)$ is a Legendre polynomial and the prime  affixed to the sum in~\eqref{kluk} means the restriction 
that $l$ is an odd, positive integer.
In this equation we have introduced the geodesic distance $\psi_{M_0 M} = \arccos (\mathbf{z}_0 \cdot \mathbf{z}) $
between the two points $M$ and $M_0$  on the unit sphere $\mathcal{S}_{2}$. The first term in the \textit{r.h.s.} of
Eq.~\eqref{Sol_S2} is thus indeed the potential created at a source-free
point $M$ by a point bi-charge located at the north pole $\mathcal{N}$.

It is the place to introduce  Dirac's function on a sphere of radius $R \neq 1$; it will be denoted
$\delta (M_0,M) = R^{-2} \delta_{\mathcal{S}_{2}}(\mathbf{z}_0, \mathbf{z}) $. Since the Laplacian
also scales as $R^{-2}$ with the radius of the sphere the potential $V_{q, M_0}^{\mathrm{bi}}(M)$ is independent
of the latter. Thus, in $\mathcal{S}_2(0,R)$, Poisson's equation for a bi-charge takes the form
\begin{equation}
\label{Pois_S2_R}
 \Delta_{\mathcal{S}_{2}(O,R)} V_{q, M_0}^{\mathrm{bi}}(M) = -2 \pi q  \left\{   \delta(M_0, M) -  \delta(\overline{M}_0, M)  \right \} \; ,
\end{equation}

The electric field given by
\begin{align}
 \boldsymbol{E}_{q, M_0}^{\mathrm{bi}}(M)& =-\frac{1}{R} \nabla_{\mathcal{S}_{2}} V_{q, M_0}^{\mathrm{bi}}(M) \nonumber \\
&=\dfrac{q}{R} \frac{1}{ \sin \psi_{M_0 M} }\mathbf{t}_{M_0 M}(M) \; ,
\end{align}
where $ \mathbf{t}_{M_0 M}(M) = \cot \psi_{M_0 M} \mathbf{z} -  \mathbf{z}_0/\sin \psi_{M_0 M} $ 
denotes the unit vector, tangent to the geodesics $M_0 M$ at point $M$, and  orientated from  point $M_0$
towards point $M$ \cite{these,Orsay_1}. Note that this vector differs from 
$ \mathbf{t}_{M_0 M}(M_0) =- \cot \psi_{M_0 M} \mathbf{z}_0 +  \mathbf{z}/\sin \psi_{M_0 M} $ which is
 the unit vector, tangent to the geodesics $M_0 M$ at point $M_0$ (also  orientated from   $M_0$
to point $M$).
One checks that the electric field $ \boldsymbol{E}_{q, M_0}^{\mathrm{bi}}(M)$ satisfies to Gauss theorem \cite{these,Orsay_1}.

Let  us consider now a \textit{pseudo-charge} made of a point charge $+q$ located at at point $M_0$ of $\mathcal{S}_{2}$ and
a uniform neutralizing background of charge density $-q/(4\pi)$.
The potential $V_{q, M_0}^{\mathrm{ps}}(M)$
at point $M$ is the solution of Poisson's equation
\begin{equation}
\label{Pois_S3}
 \Delta_{\mathcal{S}_{2}} V_{q, M_0}^{\mathrm{ps}}(M) = -2 \pi q  \{   \delta_{\mathcal{S}_{2}}(\mathbf{z}_0, \mathbf{z})
                                                                        -  \frac{1}{4 \pi}  \} \; .
\end{equation}
Eq.~\eqref{Pois_S3} can also be solved by expanding both sides on the complete basis set of spherical harmonics
with the result
\begin{align}
\label{kluk_b}
 V_{q, M_0}^{\mathrm{ps}}(M) &= \frac{q}{2} \;  {\sum_{l=1}^{\infty}} \;  \frac{2 l +1}{l(l+1)} P_l(\cos \psi_{M_0 M}) \nonumber \\
                                                 &= - \frac{q}{2} -q \log \sin \frac{\psi_{M_0 M}}{2} \; ,
\end{align}
from which the electric field is readily obtained
\begin{align}
 \boldsymbol{E}_{q, M_0}^{\mathrm{ps}}(M)& =-\frac{1}{R} \nabla_{\mathcal{S}_{2}} V_{q, M_0}^{\mathrm{ps}}(M) \nonumber \\
&=\dfrac{q}{2 R}  \cot  \frac{\psi_{M_0 M}}{2}  \; \mathbf{t}_{M_0 M}(M) \; ,
\end{align}
an expression which can also be obtained by applying Gauss theorem \cite{these,Orsay_1}.
Some remarks are in order.
\begin{itemize}
 \item{(i)} One checks that, of course one has : $V_{q, M_0}^{\mathrm{bi}}(M) = V_{q, M_0}^{\mathrm{ps}}(M) 
                                                                                         -  V_{q, \overline{M}_0}^{\mathrm{ps}}(M) $, 
         since the backgrounds of the two pseudo-charges cancell each other.
 \item{(i)} For $R\neq 1$  one has 
\begin{equation}
\label{Pois_S3_R}
 \Delta_{\mathcal{S}_{2}(O,R)} V_{q, M_0}^{\mathrm{ps}}(M) = -2 \pi q  \{   \delta(M_0,M)
                                                                        -  \frac{1}{S}  \} \; .
\end{equation}
where $S=4 \pi R^2$ is the surface of $\mathcal{S}_2(0,R)$.
\item{(iii)} Since  the $3D$ distance between points $M_0$ and $M$, \textit{i.e.} the length of the chord
        joining the two points is $d=2 R \sin ( \psi_{M_0 M}/2) $, it turns out that the expression of the 
        potential of a pseudo-charge in  $\mathcal{S}_{2}$ coincides
        with  that of a point charge in the plane $E_2$.
\item{(iv)} Although the potential of a pseudo-charge does not satisfies to Laplace-Beltrami equation at source-free
               points, the potentials created by neutral multipoles made of pseudo-charges
                indeed do,  since the backgrounds  cancell, as it will be seen below.
\end{itemize}

In order to extend the Green's function expansion~\eqref{Expand1} to the sphere we have found the following
trick. Let us introduce, as in Ref.~\cite{first}, the Cayley-Klein parameters
\begin{subequations}
\label{Cayley}
 \begin{align}
  \alpha &= \exp(i \varphi/2) \cos \frac{\theta}{2} \; , \\
 \beta   &= -i \exp(- i \varphi/2) \sin \frac{\theta}{2} \; .
 \end{align}
\end{subequations}
A short calculation reveals that, for two unit vectors $\mathbf{z}_1$ and $\mathbf{z}_2$ of
$\mathcal{S}_2$, one has
\begin{equation}
\arrowvert \alpha_1 \beta_2 - \alpha_2 \beta_1 \arrowvert = \sin  \frac{\psi_{12}}{2} \; ,
\end{equation}
from which it follows  that 
\begin{equation}
 \log  \sin  \frac{\psi_{12}}{2}  = \log \sin \frac{\theta_>}{2} +\log \cos \frac{\theta_<}{2} + \log 
  \arrowvert 1 -z \exp(i \Delta \varphi)
\arrowvert  \; ,
\end{equation}
where  $\theta_>=\sup(\theta_1,\theta_2)$, $\theta_<=\inf(\theta_1,\theta_2)$, $\Delta \varphi= \pm (\varphi_1 - \varphi_2)$, 
and  $z=\tan(\theta_</2)/\tan(\theta_>/2)$. We note that 
$ \log \arrowvert 1 -z \exp(i \Delta \varphi)  \arrowvert  = (1/2)[ \log (1 - z \exp(i \Delta \varphi)) + \log (1 - z \exp(- i \Delta \varphi)) ]$ and that, 
since $\arrowvert z \arrowvert <1$,
we can use the expansion of  the complex logarithm
$\log (1-Z)$ within its circle of convergence, \textit{i.e.}
 $\log (1-Z) = - \sum_{n=1}^{\infty}Z^n/n$
 for $\vert Z \vert < 1$, where  $Z=z \exp(\pm i \Delta \varphi)$.
In doing so,  we obtain  the following  expansion  :
\begin{align}
\label{toto}
- \log  \sin \frac{\psi_{12}}{2} &= -  \log  \sin  \frac{\theta_>}{2} -  \log  \cos  \frac{\theta_<}{2} \nonumber \\
                                               & + \sum_{n=1}^{\infty} \dfrac{1}{n}\left[ \dfrac{\tan(\theta_< /2 )}{\tan ( \theta_>/2) }\right]^n \cos (n \Delta \varphi) \; ,
\end{align}
which seems to be a new mathematical result.
In order to make something usefull of that esthetic formula, we consider now a set of $N$ pseudo-charges 
  $q_i$  of $\mathcal{S}_2(O,R)$,
with spherical coordinates $(\theta_i,\varphi_i)$, and all  around the north pole $\mathcal{N}$, \textit{i.e.}
$ \pi \geq \theta_0 >\theta_i$, $\forall i = 1, \ldots, N$.
A multipolar expansion of the electrostatic potential   $V_{\mathcal{S}_2(O,R}^{\mathrm{ps}}(M)$ 
created by these charges at some point $M = (\theta,\varphi)$ of the surface of the sphere away from $\mathcal{N}$
follows from  Eq.~\eqref{toto} under the assumption 
that $\pi \geq \theta \geq \theta_0$. One finds 
\begin{align}
\label{multi2}
 V_{\mathcal{S}_2(O,R)}^{\mathrm{ps}}(M) &= -\sum_{i=1}^N  \frac{q_i}{ 2 }-\sum_{i=1}^N q_i \log  \sin  \frac{\psi_{ M_i M}}{2}  \nonumber \\
                    &= V_0^{\mathrm{ps}}  -\frac{1}{2} \sum_{\nu}Q_ {0 \nu} \log  \left[ 2R \sin \frac{\theta}{2}  \right] +\frac{1}{2}\sum_{n=1}^{\infty} \sum_{\nu}
 \frac{1}{n X^n} \; Q_ {n \nu} \exp(-i n \nu \varphi) \; ,
\end{align}
where $V_0^{\mathrm{ps}}$ is some unessential constant. In~\eqref{multi2} the index $\nu $ assumes, as in Sec.~\ref{plane},
the values $-1$ and $+1$ only. On the sphere, the complex
multipolar moments are  defined as
\begin{equation}
 \label{Q2}
Q_ {n \nu} =  \sum_{i=1}^N q_i X_i^n \exp(i n \nu \varphi_i)  \; .
\end{equation}
In Eqs.~\eqref{multi2} and~\eqref{Q2}  we have introduced the variables  $X=2R \tan (\theta/2)$ not to be confused
with the length   $2R \sin(\theta/2)$ of the chord $\arc{NM}$  which constitutes  the argument of the ``$\log$''  term in the \textit{r.h.s.} of~\eqref{multi2}.

We note that, since $Q_ {n,- \nu}=Q_ {n \nu}^{*}$,  there are two independent
multipoles of order $n$ only (except for the charge of the distribution, \textit{i.e.} the degenerate case $n=0$).
In the thermodynamic limit $\rho = R \theta$ fixed, $R\to \infty$
we have $X \to \rho$ and the multipole moments~\eqref{Q2} as well as the 
multipolar expansion of the potential in~\eqref{multi2}
reduces to their ``Euclidian'' expressions in the plane $\mathcal{T}(\mathcal{N})$ tangent to the sphere
at the north pole $\mathcal{N}$, resp.~Eqs.~\eqref{Q1} and~\eqref{multi1}.  We remark that quite generally
a point multipole
located at some point $M$ of $\mathcal{S}_2$ coincides with a $2D$ Euclidian point multipole
in the tangent plane $\mathcal{T}(M)$ with the same complex  moments $Q_ {n \nu}$.
Therefore
the electrostatic potential  $Q_ {n \nu}\exp(-i n \nu \varphi)/(n X^n)$ in the \textit{r.h.s.} of~\eqref{multi2}
may be interpreted  as
the potential created by a   multipole of order $n$ and value $Q_ {n \nu}$ located at the north pole
$\mathcal{N}$ of the sphere. 
If $n \neq 0$   it clearly is   a non-isotropic solutions of Laplace-Beltrami equation, singular
at $\theta=0$, \textit{cf} Eq.~\eqref{Sol_S2}.

For the sake of completeness we consider now a set of $N$ bi-charges of $\mathcal{S}_2(O,R)$.
Let  $q_i$ ($i =1, \ldots, N$)  be the point charges located in 
the northern hemisphere at the points $M_i$ with spherical coordinates  $(\theta_i, \varphi_i)$. 
Their $N$ companions $-q_i$ are located
at the antipodal points    $\overline{M}_i$ in the southern-hemisphere, with coordinates $ \overline{\theta}_i= \pi -\theta_i $ and
 $\overline{\varphi}_i= \pi + \varphi_i $.
We will denote by $V_{\mathcal{S}_{2}(O,R)}^{\mathrm{bi}}(M)$ the electrostatic potential at some point $M$
of the surface of the sphere.
Assuming that $ \theta_i  < \theta  < \pi - \theta_i $, $ \forall i$ and making use of~\eqref{toto} one finds

\begin{align}
 \label{multi3}
V_{\mathcal{S}_{2}(O,R)}^{\mathrm{bi}}(M) &= 
 V_{\mathcal{S}_2(O,R)}^{\mathrm{ps}}(M)   +\sum_{i=1}^N  \frac{q_i}{ 2 }
+\sum_{i=1}^N q_i \log  \sin ( \frac{\psi_{\overline{M}_i M}}{2})  \; , \nonumber \\
               &= V_0^{\mathrm{bi}}
 -\frac{1}{2} \sum_{\nu}Q_ {0 \nu}\log  \left[ 2R \tan \frac{\theta}{2}  \right] + \nonumber \\
 &+  \frac{1}{2}\sum_{n=1}^{\infty} \sum_{\nu}
\dfrac{Q_ {n \nu} \exp(-i n \nu \varphi)  }{n [2 R]^n}
\left\{  \left[\cot \frac{\theta}{2} \right]^n  - (-1)^n \left[\tan \frac{\theta}{2} \right]^n
\right\}
\; ,
\end{align}
where $ V_0^{\mathrm{bi}}$ is some irrelevant constant. The complex
multipole moment $Q_ {n \nu} $ has been defined in Eq.~\eqref{Q2}. The contribution $n=0$
is the potential  of a point bi-charge of total charge $\sum_i q_i$ and located at the north pole $\mathcal{N}$.
The contribution  $n \neq 0$ is that of a point multipole of order $n$ and
complex moment  $Q_ {n \nu} $ located at $\mathcal{N}$
(and its dumbell companion at the south pole $\mathcal{S}$).
 Its electrostatic
potential has two singularities in $\theta=0$ and $\theta=\pi$ (as
expected) and it is one of the non-isotropic solution of Laplace-Beltrami Eq.~\eqref{Sol_S2}.
To distinguish the two types of multipoles encountered in this section, we shall
christen  mono-multipoles those obtained from pseudo-charges and  bi-multipoles
those obtained from bi-charges.

\section{Dipoles}
\label{Dipoles}
\subsection{The electric potential of a point dipole}
\label{elec-pot}
Henceforth we specialize in  the case of  dipoles. We start with a point mono-dipole
located at the north pole $\mathcal{N}$. It may be seen as a system of $N=2$ pseudo-charges of
$\mathcal{S}_2(O,R)$ : a first charge  $ \delta q$ with spherical coordinates $(\delta \theta_0 , \varphi_0)$
and a second one with opposite sign $ -\delta q$ and coordinates  $(\delta \theta_0,  \varphi _0+ \pi)$. 
We then take the limit
$\delta q \to \infty$ and $\delta \theta_0 \to 0$ with the constraint that the \textit{dipole moment}
 $\mu = 2 R \theta_0 \delta q $ is  fixed. In this limit the vectorial
 moment $\boldsymbol{\mu}=\mu \, (\cos \varphi_0 \, \mathbf{e}_x
+ \sin \varphi_0 \, \mathbf{e}_y)$ belongs to the horizontal
plane $\mathcal{T}(\mathcal{N}) \equiv ( \mathbf{e}_x, \mathbf{e}_y)$
and its two non-zero complex multipole moments are $Q_{n \nu}= \delta_{1,n} \mu \exp(i \nu \varphi_0)$
where $\nu = \pm 1$. By inserting
this expression in Eq.~\eqref{multi2} one obtains the dipolar potential at some point $M$ of the sphere 
\begin{equation}
\label{dip_mono_1}
 V_{\mathcal{N},\boldsymbol{\mu} }^{\mathrm{mono}}(M) = \frac{\mu}{2R} \cos (\varphi - \varphi_0) \cot \frac{\theta}{2} \; .
\end{equation}
In a similar way, one obtains for a bi-dipole located at the north pole $\mathcal{N}$
\begin{equation}
\label{dip_bi_1}
 V_{\mathcal{N},\boldsymbol{\mu} }^{\mathrm{bi}}(M) = \frac{\mu}{2R} \cos (\varphi - \varphi_0)
\left\{ \cot \frac{\theta}{2} + \tan \frac{\theta}{2} \right\}  \; .
\end{equation}
Note that  $V_{\mathcal{N},\boldsymbol{\mu} }^{\mathrm{mono}}(M)$ is a solution of Laplace-Beltrami Eq.~\eqref{Sol_S2} 
since the backgrounds of the two charges $\pm \delta q$ cancell each other. There is a single singularity
in $\theta=0$. 	In the thermodynamic limit  $R \to \infty$, $\rho=R \theta$ fixed one recover the Euclidian dipolar
potential $\boldsymbol{\mu} \cdot\boldsymbol{\rho} / \rho^2 $ of a $2D$ dipole of the $( \mathbf{e}_x, \mathbf{e}_y)$ plane.
$V_{\mathcal{N},\boldsymbol{\mu} }^{\mathrm{bi}}(M)$ is also a solution of Laplace-Beltrami Eq.~\eqref{Sol_S2} but
with two singularities at   $\theta=0$ and $\theta=\pi$. Note that the dipoles at the south-pole and the north-pole
are the same vectors of $E_3$. The additional term $\tan (\theta /2)$ in the potential  vanishes in the thermodynamic
limit $R \to \infty$, $\rho=R \theta$ fixed, and does not change the conclusion that one  also recovers
the  Euclidian dipolar  potential in this limit for the bi-dipole.

In the general case where the dipole say $\boldsymbol{\mu}_0  $ of modulus $\mu = \|  \boldsymbol{\mu}_0    \| $
is located at some point $M_0$ of 
$\mathcal{S}_2(O,R)$ one obviously have
\begin{subequations}
\begin{align}
\label{pot_dip}
 V_{M_0,\boldsymbol{\mu}_0 }^{\mathrm{mono}}(M) &= \frac{\mu}{2R} \, \cot \frac{\psi_{M_0 M}}{2} \,
                         \mathbf{s}_0 \cdot \mathbf{t}_{M M_0}(M_0)  \nonumber \\
&  =\frac{\mu}{4R} \dfrac{1 }{\sin^2 \frac{\psi_{M_0 M}}{2}} \; \mathbf{s}_0 \cdot \mathbf{z} 
  \\
 V_{M_0,\boldsymbol{\mu}_0 }^{\mathrm{bi}}(M) &= V_{M_0,\boldsymbol{\mu}_0 }^{\mathrm{mono}}(M) +
                                                          V_{\overline{M}_0,\boldsymbol{\mu}_0 }^{\mathrm{mono}}(M)  \label{Strontium}\\
 &= \frac{\mu}{R} \,
\dfrac{ 1}{\sin \psi_{M_0 M}} \, \mathbf{s}_0 \cdot \mathbf{t}_{M M_0}(M_0) \nonumber \\
& =\frac{\mu}{R} \dfrac{ 1 }{\sin^2 \psi_{M_0 M} } \;  \mathbf{s}_0 \cdot \mathbf{z} 
 \; ,
\end{align}
\end{subequations}
where we have introduced the unit vector $\mathbf{s}_0= \boldsymbol{\mu_0}/ \mu$.
\subsection{The electric field of a point dipole}
\label{elec-field}
The electric field created by a mono-dipole  located at some point $M_0$ is obtained by taking minus the gradient of 
the potential $V_{M_0, \boldsymbol{\mu}_0}^{\mathrm{mono}}(M)$ at point $M$ with the result :
\begin{equation}
\label{field}
 \mathbf{E}_{M_0, \boldsymbol{\mu}_0}^{\mathrm{mono}}(M) = -
 \dfrac{1}{R} \cdot \nabla_{\mathcal{S}_3, M}  V_{M_0, \boldsymbol{\mu}_0 }^{\mathrm{mono}}(M)  = 2 \pi
\mathbf{G}_0^{\mathrm{mono}} (M, M_0) \cdot \boldsymbol{\mu}_0\; ,
\end{equation}
where we have introduced, as in Ref.~\cite{Tr_Cai}, the tensorial dipolar Green's function $\mathbf{G}_0^{\mathrm{mono}} (M,
M_0) $, for which we give two useful expressions :
\begin{subequations}
\label{G0_mono}
  \begin{align}
\mathbf{G}_0^{\mathrm{mono}} (M,M_0) =& \frac{1}{4 \pi R^2}
                                                                        \frac{1}{1 - \cos \psi_{M_0M}}  \, \nonumber \\
 &\left[ \left(1+  \cos \psi_{M_0M} \right)  \mathbf{t}_{M M_0} (M)\mathbf{t}_{M M_0} (M_0) - \label{G0_mono_a} 
\mathbf{U}_{\mathcal{S}_2}(\mathbf{z}) \cdot  \mathbf{U}_{\mathcal{S}_3}(\mathbf{z}_0)\right] 
  \; ,    \\
=& - \frac{1}{R^2} \sum_{l=1}^{\infty}  \sum_{m=-l}^{+l} \frac{1}{l(l+1)} \nabla_{\mathcal{S}_3}\overline{Y}^{m }_{l}(\mathbf{z})
          \nabla_{\mathcal{S}_3}  Y_{l}^{m}(\mathbf{z}_0)  \; , \label{G0_mono_b} 
 \end{align}
\end{subequations}
as a short algebra will show.  
We stress that $\mathbf{G}_0^{\mathrm{mono}} (M,M_0)$ is a $3D$ dyadic
tensor of the type $\mathbf{A}(M) \mathbf{A}(M_0)$, $\mathbf{A}(M)$ and $ \mathbf{A}(M_0)$ being
two vectors tangent to the sphere at the points $M$ and $M_0$, respectively. It is easy to show that
in the limit $\psi_{M_0M} \to 0$, $\mathbf{G}_0 ^{\mathrm{mono}}(M,M_0)$ tends to its Euclidian limit
$\mathbf{G}_{0, E_2} (M,M_0)=  [-\mathbf{U}_{E_2}(M) + 2 \widehat{\boldsymbol{\rho}}
 \widehat{\boldsymbol{\rho}} ] /(2 \pi \rho^2)$, with  $\boldsymbol{\rho} = \overrightarrow{M_0M}$ 
and $\widehat{\boldsymbol{\rho}}=  \boldsymbol{\rho} /\rho$ and where $\mathbf{U}_{E_2}(M)
= \mathbf{e}_{\theta}\mathbf{e}_{\theta} +\mathbf{e}_{\varphi}\mathbf{e}_{\varphi}$ is the unit dyadic
tensor in the tangent plane $\mathcal{T}(M)$.

With arguments similar to those exposed in Ref.~\cite{Jackson,Fulton_I,Fulton_II,Fulton_III} for the $3D$ case,
one can easily show that the distribution $\mathbf{G}_{0, E_2} (M,M_0)$ has a singularity
$-(1/2) \mathbf{U} \delta^{(2)}(\boldsymbol{\rho} )$.  Therefore
 $\mathbf{G}_0 ^{\mathrm{mono}} (M,M_0)$ is singular
for $\psi_{M_0M} \to 0$, with the same singularity. As for the $3D$ case~\cite{Tr_Cai} it may be important to extract 
this singularity and to define
a non-singular Green's  function  $\mathbf{G}_0^{\mathrm{mono},  \delta} (M,M_0)$ by the relations
\begin{subequations}
 \begin{align} \label{decompo} 
 \mathbf{G}_0^{\mathrm{mono}}(M,M_0) & =   \mathbf{G}_0^{\mathrm{mono},  \delta} (M, M_0) - \, \frac{1}{2}\delta(M,M_0) \,     
                                             \mathbf{U}_{\mathcal{S}_2}(\mathbf{z}) \; ,  & \\
\mathbf{G}_0^{\mathrm{mono}, \delta} (M,M_0)  &= \begin{cases}
                                                  \mathbf{G}_0^{\mathrm{mono}}(M,M_0)\; &,   \text{ for } R  \psi_{M_0M} > \delta  \; ,  \\        
                                                  0                        \;  &,    \text{ for } R  \psi_{M_0M} < \delta  \; , 
                                                     \end{cases}
 \end{align}
\end{subequations}
where $\delta$ is an arbitrary small cut-off ultimately set to zero. It must be understood that any integral
involving $\mathbf{G}_0^{\mathrm{mono},  \delta} $ must be calculated with $\delta \neq 0$
and then taking the limit $\delta \to 0$. These matters are discussed in  appendix~\ref{Appendix_A} together with
some other useful mathematical properties of $\mathbf{G}_0^{\mathrm{mono}} (M,M_0)$.

For the bi-dipoles one finds
\begin{subequations}
\label{G0_bi}
  \begin{align}
\mathbf{G}_0^{\mathrm{bi}} (M,M_0) = &\mathbf{G}_0^{\mathrm{mono}} (M,M_0)  +
 \mathbf{G}_0^{\mathrm{mono}} (M,\overline{M}_0) \nonumber \\
=& \frac{1}{2 \pi R^2}
                                                                        \frac{1}{ \sin^{2} \psi_{M_0M}}  \, \label{Sodium} \\
 &\left[ 2 \cos \psi_{M_0M}  \mathbf{t}_{M M_0} (M)\mathbf{t}_{M M_0} (M_0) - \label{G0_bi_a} 
\mathbf{U}_{\mathcal{S}_2}(\mathbf{z}) \cdot  \mathbf{U}_{\mathcal{S}_3}(\mathbf{z}_0)\right] 
  \; ,    \\
=& - \frac{2}{R^2} \sum_{l \; \text{odd}}  \sum_{m=-l}^{+l} \frac{1}{l(l+1)} \nabla_{\mathcal{S}_3}\overline{Y}^{m }_{l}(\mathbf{z})
          \nabla_{\mathcal{S}_3}  Y_{l}^{m}(\mathbf{z}_0)  \; . \label{G0_bi_b}
 \end{align}
\end{subequations}
in addition to a singularity for $M \to M_0$ (the same as that of $\mathbf{G}_0^{\mathrm{mono}} (M,M_0)$)
the Green's function $\mathbf{G}_0^{\mathrm{bi}} (M,M_0)$ bears another singularity 
for  $M \to \overline{M_0}$.

\subsection{Dipole-dipole interaction}
\label{dip_dip}
We end this section by \textit{defining } the interaction of two dipoles $(M_1,\boldsymbol{\mu}_1)$
and $(M_2,\boldsymbol{\mu}_2)$ as $W_{\boldsymbol{\mu}_1,  \boldsymbol{\mu}_2} \equiv - \boldsymbol{\mu}_1 \cdot
2 \pi  \mathbf{G}_0(1,2) \cdot \boldsymbol{\mu}_2 $ which gives for the mono dipoles with the help of Eq.~\eqref{G0_mono_a}

\begin{subequations}
\label{ene_mono}
\begin{align}
 W_{\boldsymbol{\mu}_1 ,\boldsymbol{\mu}_2}^{\mathrm{mono}} &= \frac{\mu^2}{2 R^2}\frac{1}{1 - \cos \psi_{12}}
\bigg(  \mathbf{s}_1  \cdot  \mathbf{s}_2    
 -    (1+ \cos \psi_{12})\, ( \mathbf{t}_{12}(1) \cdot    \mathbf{s}_1)
  ( \mathbf{t}_{12}(2) \cdot   \mathbf{s}_2 ) \bigg)  \; ,  \\
&= \frac{\mu^2}{2R^2}\frac{1}{1 - \cos \psi_{12}} \bigg(   \mathbf{s}_1  \cdot  \mathbf{s}_2  
+ \frac{1}{1- \cos \psi_{12}  } \, (   \mathbf{s}_1 \cdot \mathbf{z}_2)
 ( \mathbf{s}_2 \cdot\mathbf{z}_1)
 \bigg) \; , \label{dip_mono}
\end{align}
\end{subequations}
were we have adopted the simplified notation $\mathbf{t}_{12}(i) \equiv \mathbf{t}_{M_1 M_2}(M_i) \; (i=1,2)$
to denote the two unit vectors tangent to the
geodesic  $\arc{M_1 M_2}$.
The interaction energy of two bi-dipoles is given by
\begin{subequations}
\label{ene_bi}
\begin{align}
 W_{\boldsymbol{\mu}_1 ,\boldsymbol{\mu}_2}^{\mathrm{bi}} &= \frac{\mu^2}{R^2}\frac{1}{\sin^2 \psi_{12}}
\bigg(  \mathbf{s}_1  \cdot  \mathbf{s}_2    
 - 2 \cos \psi_{12}\, ( \mathbf{t}_{12}(1) \cdot    \mathbf{s}_1 )
  ( \mathbf{t}_{12}(2) \cdot   \mathbf{s}_2 ) \bigg)  \; ,  \\
&= \frac{\mu^2}{R^2}\frac{1}{\sin^2 \psi_{12}} \bigg(   \mathbf{s}_1  \cdot  \mathbf{s}_2  
+  \frac{2 \cos \psi_{12}}{\sin^2 \psi_{12}}\, (   \mathbf{s}_1 \cdot \mathbf{z}_2)
 ( \mathbf{s}_2 \cdot\mathbf{z}_1)
 \bigg) \; .  \label{dip_bi}
\end{align}
\end{subequations}
One recovers the well-known Euclidian limit
$ W_{\boldsymbol{\mu}_1 ,\boldsymbol{\mu}_2}^{\mathrm{mono}} \sim
W_{\boldsymbol{\mu}_1 ,\boldsymbol{\mu}_2}^{\mathrm{bi}}  \sim (\mu^2 /\rho_{12}^2) [\mathbf{s}_1 \cdot \mathbf{s}_2
-2( \mathbf{s}_1  \cdot \widehat{\boldsymbol{\rho}}_{12})  ( \mathbf{s}_2  \cdot \widehat{\boldsymbol{\rho}}_{12})   ]$
of the dipole-dipole interaction when \mbox{$\psi_{12} \to 0$}.
\section{The $2D$ polar fluid in $\mathcal{S}_2(O,R)$ .}
\label{2D-Polar-Fluid}

\subsection{Two equivalent models for the dipolar hard sphere fluid in  $\mathcal{S}_2(O,R)$}
\label{DHS}
A fluid of point dipoles is not stable and additional repulsive short range pair-potentials
must be included in the model to ensure the existence of a thermodynamic limit. In this paper we consider
only  hard core repulsions and thus the DHS model. Two versions are possible in   $\mathcal{S}_2(O,R)$
depending on whether mono- or bi-dipoles are involved \cite{Tr_Cai}.
\subsubsection{Monodipoles}
\label{Mono-dipoles}
In this version the mono-dipoles are embedded in the center of  hard disks 
 lying on the surface of the sphere.
In a given configuration, $N$   point-dipoles $\boldsymbol{\mu}_i$ are  located at  points
$\mathbf{OM}_i=R \mathbf{z}_i$  ($i=1, \ldots,N$) 
of $\mathcal{S}_2(O,R)$ and  the configurational energy reads
\begin{equation}
\label{ }
 U (\{\mathbf{z}_i, \boldsymbol{\mu}_i \})=  \frac{1}{2} \sum_{i \neq j}^N \; v_{\mathrm{HS}}^{\mathrm{mono}}(\psi_ {ij}) + \frac{1}{2}
\sum_{i \neq j}^N \;  W_{\boldsymbol{\mu}_i, \boldsymbol{\mu}_j}^{\mathrm{mono}}  \; ,
\end{equation}
where $v_{\mathrm{HS}}^{\mathrm{mono}}(\psi_ {ij}) $ is the hard-core pair potential in $\mathcal{S}_2(O,R)$
defined by
\begin{equation}
 v_{\mathrm{HS}}^{\mathrm{mono}}(\psi_ {ij}) =  \begin{cases}
                                                 \infty & \text{ if }  \sigma/R > \psi_{ij}   \; ,   \\
                                                 0&       \text{ otherwise } \; ,  
                                                \end{cases}
\end{equation}
and $ W_{\boldsymbol{\mu}_i, \boldsymbol{\mu}_j}^{\mathrm{mono}}$ is the energy of a pair of mono-dipoles
given at Eq.~\eqref{ene_mono}. Note that the distance $\sigma$ is measured along the geodesics
and not in the Euclidian space $E_3$. The hard disks are in fact curved objects, similar to contact
lenses at the surface of an eye-ball. 

A thermodynamic state of this model is characterized by a density $\rho^*=N\sigma^2/S$ where $S=4 \pi R^2$
is the $2D$ surface  of the sphere $\mathcal{S}_2(O,R)$ and a reduced dipole moment
$\mu^*$ with  $\mu^{*2}= \mu^2/(k_B T\sigma^2) $ ($k_B $ Boltzmann constant, $T$ absolute temperature).
\subsubsection{Bi-dipoles}
In the second version we consider  a fluid of dipolar dumbells
confined on the surface of the sphere $\mathcal{S}_2(O,R)$.
Both dipoles of the dumbell are embedded at the center of a hard sphere of diameter
$\sigma$.
In a given configuration,
$N$ dipoles $\boldsymbol{\mu}_i$ are located at  points $\mathbf{OM}_i=R \mathbf{z}_i$ 
and their $N$ companions $\boldsymbol{\mu}_i$ at the antipodal points $\mathbf{O\overline{M}}_i= - R \mathbf{z}_i $
($i=1, \ldots,N$) of $\mathcal{S}_2(O,R)$. Each pair constitute a bi-dipole and only
one of the two dipoles lies in the Nothern hemisphere  $\mathcal{S}_2(O,R)^{+}$.  The configurational energy reads 
\begin{equation}
\label{conf2}
 U(\{\mathbf{z}_i, \boldsymbol{\mu}_i \}) =  \frac{1}{2} \sum_{i \neq j}^N \; v_{\mathrm{HS}}^{\mathrm{bi}}(\psi_ {ij}) + \frac{1}{2}
\sum_{i \neq j}^N \;  W_{\boldsymbol{\mu}_i, \boldsymbol{\mu}_j}^{\mathrm{bi}}  \; ,
\end{equation}
where $v_{\mathrm{HS}}^{\mathrm{bi}}(\psi_ {ij}) $ is  hard-core pair potential  defined by
\begin{equation}
 v_{\mathrm{HS}}^{\mathrm{bi}}(\psi_ {ij}) =  \begin{cases}
                                                 \infty & \text{ if }  \sigma/R > \psi_{ij}  \text{ or }    \psi_{ij}> \pi -\sigma/R  \; ,   \\
                                                 0&       \text{ otherwise }  \; ,  
                                                \end{cases}
\end{equation}
and the dipole-dipole interaction $W_{\boldsymbol{\mu}_i, \boldsymbol{\mu}_j}^{\mathrm{bi}} $ is 
that defined at Eq.~\eqref{ene_bi}. In Eq.~\eqref{conf2} the vectors $\mathbf{z}_i$ can allways be chosen
in the northern hemisphere  $\mathcal{S}_2(O,R)^+$ because of the special symmetries of the interaction.
Indeed it we have by construction $ V_{M_i,\boldsymbol{\mu}_i }^{\mathrm{bi}}(M) =
 V_{\overline{M}_i,\boldsymbol{\mu}_i }^{\mathrm{bi}}(M)$ (cf Eq.~\eqref{Strontium})
and $ \mathbf{E}_{M_i,\boldsymbol{\mu}_i }^{\mathrm{bi}}(M) =
 \mathbf{E}_{\overline{M}_i,\boldsymbol{\mu}_i }^{\mathrm{bi}}(M)$ (cf Eq.~\eqref{Sodium}) for any point
$M$ of $\mathcal{S}_2(O,R)^+$  and any configuration of bi-dipoles.
It is thus clear that the genuine  domain
occupied by the fluid is  the northern hemisphere  $\mathcal{S}_2(O,R)^+$ rather than the whole
hypersphere.
The interpretation of the model is therefore the  following.
When a dipole $\boldsymbol{\mu}_i$ quits $\mathcal{S}_2(O,R)^+$ at some point $M_i$ of the equator
the same $\boldsymbol{\mu}_i$  reenters at the antipodal point  $\overline{M}_i$. Therefore bi-dipoles living
on the whole sphere are equivalent to mono-dipoles living on the northern hemisphere but with
special  boundary conditions ensuring homogeneity and isotropy at equilibrium (in the case of a fluid phase). 

A thermodynamic state of this model is now characterized by a density $\rho^*=N\sigma^2/S$ where $S=2 \pi R^2$
is the $2 D$ surface  of the northern hemisphere  $\mathcal{S}_2^+(O,R)$ and the  reduced dipole
$\mu^*$ with  $\mu^{*2}= \mu^2/(k_B T\sigma^2) $ as in Sec.~\ref{Mono-dipoles}. 

\subsection{Thermodynamic and structure}
\label{Thermo}
Most properties of the DHS fluid at thermal equilibrium can be obtained from the knowledge of the
one- and two-body correlation functions  defined on the sphere and in the canonical ensemble as
\begin{subequations}
\label{correlations}
 \begin{align}
  \rho^{(1)}(1)& = \left\langle \sum_{i=1}^{N} \delta(M_1, M_i) \, \delta(\alpha_1 - \alpha_i) \right\rangle \\ 
\label{ro2}
  \rho^{(2)}(1,2)& = \left\langle \sum_{i=1}^{N}\sum_{j=1}^{N} (1- \delta_{i j}) \delta(M_1, M_i)   \, \delta(M_2, M_j) \, 
 \delta(\alpha_1 - \alpha_i) \,  \delta(\alpha_2 - \alpha_j) 
 \right\rangle \; .
 \end{align}
\end{subequations}
 Here $(i) \equiv (\mathbf{z}_i, \alpha_i)$  denotes both the position $\mathbf{z}_i$ of
dipole ``i'' and its orientation, specified by the angle $\alpha_i$. This angle can be measured for instance
in the local basis ($\mathbf{e}_{\theta_i}$,  $\mathbf{e}_{\varphi_i}$) of spherical coordinates, \textit{i.e.}
$\alpha_i =  \widehat{(\mathbf{e}_{\theta_i}, \mathbf{s}_i )}$. 
For a homogeneous and isotropic fluid one has $  \rho^{(1)}(1)=\rho/(2 \pi)$ where $\rho=N/S$ is the 
number density and
\begin{equation}
  \rho^{(2)}(1,2) = \left( \dfrac{\rho}{2 \pi} \right)^2 \, g(1,2)  \; ,
\end{equation}
where the correlation function $g(1,2)$ has been normalized so as to tend to $1$ at large distances.
To exploit the symmetries of the fluid phase It is  far more convenient to use, instead of the $\alpha_i$,
the angles $\beta_i =  \widehat{(\mathbf{t}_{12}(i), \mathbf{s}_i )} \; (i=1,2)$.
 In order to complete the  local basis
in the tangent planes $\mathcal{T}_i \, (i=1,2)$ one defines 
$\mathbf{n}_{12}(i) = \mathbf{z}_i \times \mathbf{t}_{12}(i) \; (i=1,2)$. Remarkably the
vector $\mathbf{n}_{12}$ is invariant along the geodesic $\arc{M_1 M_2}$ and one has
\begin{equation}
\label{n12}
 \mathbf{n}_{12}(i) \equiv  \mathbf{n}_{12} =\dfrac{\mathbf{z}_1 \times \mathbf{z}_2}{\sin \psi_{12}} \; .
\end{equation}
Note by passing that the unit dyadic tensor $\mathbf{U}_{\mathcal{S}_2}(\mathbf{z}_i)=\mathbf{n}_{12} \mathbf{n}_{12} +
  \mathbf{t}_{12}(i) \mathbf{t}_{12}(i) $ for $i=1,2$.

These variables $(\beta_1, \beta_2)$ serve mainly to introduce a basis of $2D$ rotational invariants 
$\Psi_{mn}(\beta_1, \beta_2)$ upon which to expand
the pair correlation function $g(1,2)$. The rotational invariants were introduced by Blum and Toruella 
in the context of $3D$ homogeneous and isotropic molecular fluids~\cite{Blum} and their $2D$ counterparts
 in the $2D$ plane were given in Refs~\cite{these,Orsay_1}. On the sphere the
extension of the latter result is straightforward and these invariants read
\begin{equation}
 \Psi_{mn}(1,2) \equiv \exp \left(i m \beta_1 +i n \beta_2 \right)   \; .
\end{equation}
These functions are indeed invariant under the action of a global rotation of $E_3$ of center $O$.
Moreover these invariants are orthogonal is the sense that 
\begin{equation}
 <\Psi_{mn} \vert \Psi_{pq}> \equiv \int \dfrac{d \beta_1}{2 \pi} \int \dfrac{d \beta_2}{2 \pi}\overline{\Psi}_{mn}(1,2)\Psi_{pq}(1,2)
=\delta_{mp}\delta_{nq} \; .
\end{equation}
and in an  homogeneous and isotropic phase of the DHS fluid one has the expansion
\begin{equation}
\label{rotas}
 g(1,2) = \sum_{m,n = - \infty}^{\infty} g_{mn}(r_{12})\Psi_{mn}(1,2)  \;,
\end{equation}
 where the coefficients $g_{mn}(r_{12})$ are functions of the length $r_{12}= R \psi_{12}$
of the geodesic $\arc{M_1 M_2}$.

Reality of the correlation function $g(1,2)$ and invariance through the reflexion
$ (\beta_1, \beta_2 )\leftrightarrow (-\beta_1,- \beta_2 )$ impose 
that $g_{mn}=g_{-m -n}=g_{mn}^* $. These conditions enable us to define the real rotational invariants
\begin{align}
 \Phi_{mn} &= \frac{1}{2} \left( \Psi_{mn} +\Psi_{-m -n}  \right) \nonumber \\
                 & = \cos \left( m \beta_1 + n \beta_2 \right) \; . 
\end{align}
To give some physical and geometrical interpretation of some of these invariants we first
recall the expression of the scalar product on the sphere $\mathcal{S}_n(O,R)$~\cite{these,Orsay_1,Orsay_2,Tr_Cai}.
Taking the scalar product $\Delta(1,2)$ of
two vectors  $\mathbf{s}_1$ and $\mathbf{s}_2$ located at two distinct points,
 $M_1$ and $M_2$ of $\mathcal{S}_2(O,R)$ needs some caution. It first requires 
to perform a parallel transport of vector   $\mathbf{s}_1$ from 
 $M_1$ to $M_2$ along the geodesic $\arc{M_1M_2}$ and then to take a $2\mathrm{D}$
scalar product in space  $\mathcal{T}(M_2)$.
Thus
\begin{equation}
\label{scal}
\Delta(1,2)  = \tau_{12} \mathbf{s}_1 \cdot \mathbf{s}_2 \; ,
\end{equation}
where, in the r.h.s. the dot denotes the usual scalar product of the Euclidian space 
\mbox{$ \mathcal{T}(M_2)$.}
Vector  $\tau_{12} \mathbf{s}_1 $  results from a  transport of $\mathbf{s}_1$ 
from the space  $\mathcal{T}(M_1)$ to  the space  $\mathcal{T}(M_2)$ along the geodesic
 $M_1M_2$, keeping its angle with the tangent to the geodesic constant. Explicitely
one has:
\begin{equation}
 \label{transport}
 \tau_{12} \mathbf{s}_1 =\mathbf{s}_1 - \frac{ \mathbf{s}_1 \cdot \mathbf{z}_2}{1+\cos \psi_{12}} \;  
\left(  \mathbf{z}_1+  \mathbf{z}_2  \right) \; ,
\end{equation}
and thus
\begin{equation}
\label{scal_bis}
 \Delta(1,2) =   \mathbf{s}_1 \cdot \mathbf{s}_2
- \frac{(  \mathbf{s}_1 \cdot \mathbf{z}_2)  (  \mathbf{s}_2 \cdot \mathbf{z}_1)    }{ 1+\cos \psi_{12} } \; .
\end{equation}

The above expression~\eqref{scal_bis} is in fact valid for a hypersphere   $\mathcal{S}_n(O,R)$ of arbitrary dimensions.
In the case $n=2$ more simple expressions can easily be obtained with the help of the angles
$\beta_i$ introduced previously. Starting
from $ \mathbf{s}_1 = \cos \beta_1 \mathbf{t}_{12}(1) + \sin \beta_1 \mathbf{n}_{12}$  one deduces
from~\eqref{transport} that $ \tau_{12} \mathbf{s}_1 = \cos \beta_1 \mathbf{t}_{12}(2) + \sin \beta_1 \mathbf{n}_{12}$ 
from which it follows that
\begin{equation}
 \Delta(1,2) = \cos(\beta_1 - \beta_2) \; ,
\end{equation}
and therefore $\Delta \equiv  \Phi_{1 -1}$.

By analogy with the $2D$ Euclidian case we also introduce the manifestly rotationally
invariant function $D(1,2)$
\begin{align}
 D(1,2)\equiv &= 2 \big(\mathbf{s}_1 \cdot \mathbf{t}_{12}(1) \big)     \big(\mathbf{s}_2 \cdot \mathbf{t}_{12}(2) \big)             - \Delta(1,2) \nonumber \\
                     &= - \mathbf{s}_1 \cdot \mathbf{s}_2
- \frac{(  \mathbf{s}_1 \cdot \mathbf{z}_2) (  \mathbf{s}_2 \cdot \mathbf{z}_1)    }{ 1-\cos \psi_{12} } \; .
\end{align}
An explicit calculation shows that in $\mathcal{S}_2(O,R)$ one also has
\begin{equation}
 D(1,2) = \cos(\beta_1 + \beta_2) \; ,
\end{equation}
and therefore $D \equiv  \Phi_{1 1}$.

The angular Dependance of the dipole-dipole interactions~\eqref{ene_mono} and ~\eqref{ene_bi}
should be rotationally invariant and indeed one finds that it is  a combination of the invariants $D$ and $\Delta$.
More precisely one has 
\begin{subequations}
 \begin{align}
   W_{\boldsymbol{\mu}_1 ,\boldsymbol{\mu}_2}^{\mathrm{mono}} &=
- \dfrac{\mu^2}{R^2 \sin^2 \psi_{12}}    \dfrac{1 + \cos  \psi_{12}}{2} D(1,2)  \\
 W_{\boldsymbol{\mu}_1 ,\boldsymbol{\mu}_2}^{\mathrm{bi}} &=
  - \dfrac{\mu^2}{R^2 \sin^2 \psi_{12}}  \left[ \left(\dfrac{1 + \cos  \psi_{12}}{2} \right) D(1,2)  -  
\left(\dfrac{1 - \cos  \psi_{12}}{2} \right) \Delta(1,2)    \right]    \; .
 \end{align}
\end{subequations}

The expressions of the projection $g^{mn}(r_{12})$ of the pair correlation $g(1,2)$ on 
the rotational invariants are easily deduced from the definition~\eqref{ro2} of 
the two point function $\rho^{(2)}(1,2)$ and the  properties of orthogonality of the $\Phi_{mn}$. One finds
(in the canonical ensemble)
\begin{equation}
\label{projo}
 g^{mn}(r_{12}) = \dfrac{1}{ \left \langle \Phi_{mn} \vert \Phi_{mn} \right \rangle } \,
                         \dfrac{1}{N \rho }
  \left\langle     \dfrac{ \sum_{i\neq j }  \Phi_{mn}(i,j)  \chi(\psi_{ij}-\psi_{12})   }{2 \pi R^2\sin(\psi_{ij})\delta \psi}             \right \rangle \, ,
\end{equation}
where $\delta \psi$ is the bin size and $\chi$ is defined as
\begin{equation}
\chi(\psi-\psi_{12}) = \left\{ \begin{array}{ll} 1 & \mbox{if $\psi_{12} < \psi < \psi_{12}+\delta \psi$} \\
0 & \mbox{otherwise.}
\end{array} \right.
\end{equation}
Note that  $ \left \langle D \vert D \right \rangle =  \left \langle \Delta \vert \Delta \right \rangle =1/2$
while  $ \left \langle \Phi^{00} \vert \Phi^{00} \right \rangle = 1$.
We want to point out that 
for the DHS fluid of mono-dipoles $0< \psi _{12}=r_{12}/R< \pi$, however,  for the fluid of bi-dipoles, only the range
 $0< \psi_{12} < \pi /2 $ is available, because of  the special boundary conditions involved in the model.

If follows from the precedent developments that, at equilibrium, the mean energy per par particles reads 
\begin{subequations}
\label{uh1}
  \begin{align}
  \beta u^{\mathrm{mono}} & \equiv \dfrac{<\sum_{i \neq j}  W_{\boldsymbol{\mu}_i ,\boldsymbol{\mu}_j}^{\mathrm{mono}} >
                   }{2N} \nonumber \\
         &=- \dfrac{y}{4} \int_0^{\pi} d \psi \,  \sin \psi   \dfrac{h^{D}(R \psi)}{1 - \cos \psi}  \, ,
  \end{align}
\end{subequations}
for mono-dipoles and 
\begin{subequations}
\label{uh2}
  \begin{align}
  \beta u^{\mathrm{bi}} & \equiv \dfrac{<\sum_{i \neq j}  W_{\boldsymbol{\mu}_i ,\boldsymbol{\mu}_j}^{\mathrm{mono}} >
                   }{2N} \nonumber \\
         &=- \dfrac{y}{4} \int_0^{\pi/2} d \psi \,  \sin \psi  
                \left  [ \dfrac{h^{D}(R \psi)}{1 - \cos \psi}  
 -                      \dfrac{h^{\Delta}(R \psi)}{1 +\cos \psi} 
  \right ] \, ,
  \end{align}
\end{subequations}
for bi-dipoles. In Eqs~\eqref{uh1} and~\eqref{uh1}   we have introduced the dimensionless parameter
$y = \pi \rho \beta \mu^2$ and $h\equiv g-1$. 

For the sake of exhaustivity we finally quote the expression of the compressibility factor
\begin{equation}
\label{Z}
   Z^{\mathrm{mono}\, (\mathrm{bi})}=1 +   \beta u^{\mathrm{mono}\, (\mathrm{bi})} + \dfrac{\pi}{2} \rho^{*}
\left[ \frac{R}{\sigma} \sin \frac{\sigma}{R} \right] g^{00}(\sigma +0) \; .
\end{equation}
Note the prefactor  of the isotropic projection
$ g^{00}(\sigma +0)$ at contact which accounts for curvature effects. The usual Euclidian expression 
$Z_{\infty}= 1 + \beta u_{\infty} + (\pi \rho^{*}/2) g^{00}(\sigma +0)$ emerges
in the limit $R \to \infty $, both for mono- and bi-dipoles, as it was expected.


\subsection{Fulton's theory}
\label{FULT}
Let us  consider quite generally a polar fluid occupying  a $2D$ surface $\Lambda$ with boundaries $\partial \Lambda$.
We assume the system at thermal equilibrium in a homogeneous and isotropic fluid phase. The fluid behaves
macroscopically as a dielectric medium characterized by a dielectric constant $\epsilon$.
Due to the lack of screening in such fluids, the asymptotic behavior of the pair correlation function is long ranged and
depends on the geometry of the system, \textit{i.e.} its shape, size, and the  properties imposed to
the electric field (or potential) on the boundaries 
$\partial \Lambda$ as well. 
As a consequence, the expression of the dielectric constant $\epsilon$ in terms of the fluctuations of polarization
also depends on the geometry.
These issues can  be formally taken into account in the framework of Fulton's theory~\cite{Fulton_I,Fulton_II,Fulton_III}
which achieves
an elegant  synthesis between  the linear response theory  and 
the electrodynamics of continuous media.

This formalism can be extended without more ado to non-euclidian geometries and was applied
notably  for $3D$ hyperspheres and cubico-periodical systems in Refs.~\cite{Caillol_4,Tr_Cai}.
 In this section  it is applied to the $2D$ sphere both for mono- and bi-dipoles
(a separate treatment of each model is  however necessary).
To our knowledge the case of the $2D$ plane $E_2$ was never considered in the framework 
of Fulton's theory and it is given a special treatment
in appendix~\eqref{Appendix_B}. 

\subsubsection{Mono-dipoles}
\label{Fult_mono}

We consider a fluid of
 $N$ mono-dipoles in  $\mathcal{S}_2(O,R)$ at thermal equilibrium in the presence 
of an external electrostatic field $\boldsymbol{\mathcal{E}}(M)\in \mathcal{T}(M) $.
The medium  acquires a macroscopic polarization 
\begin{equation}
 \mathbf{P} (M) = < \widehat{\mathbf{P}} (M)>_{\boldsymbol{\mathcal{E}}}  \; ,
\end{equation}
where the brackets denote the equilibrium average of the microscopic polarization
$\widehat{\mathbf{P}} (M)$ in the presence of the external field    $\boldsymbol{\mathcal{E}}$. 
In $\mathcal{S}_{2}(O,R)$ the microscopic polarization $\widehat{\mathbf{P}} (M)$ is defined 
 as~\cite{Caillol_4,Tr_Cai}
\begin{equation}
\label{micro_P}
\widehat{\mathbf{P}} (M) = \sum_{j=1}^N   \mathbf{U}_{\mathcal{S}_2}(\mathbf{z})  \cdot
\boldsymbol{\mu}_j \, \delta(M,M_j) \; ,
\end{equation}
where the unit dyadic tensor $\mathbf{U}_{\mathcal{S}_2}(\mathbf{z})$ in the r.h.s. of Eq.~\eqref{micro_P}
ensures that vector $\widehat{\mathbf{P}} (M)$
belongs to the plane $\mathcal{T}(M)$ tangent to the sphere at point $M$.

The relation between the macroscopic polarization $\mathbf{P}(M)$ and the external field 
 $\boldsymbol{\mathcal{E}}(\mathbf{r}) $ can be established in the framework of
linear-response theory, provided that  $\boldsymbol{\mathcal{E}}(M)$
is small enough,  with the result
\begin{equation}
\label{LRT}
 2 \pi \mathbf{P}(M) = \boldsymbol{\chi} \circ  \boldsymbol{\mathcal{E}} \left(\equiv 
 \int_{\mathcal{S}_{2}(O,R)}  d S^{'} \; \boldsymbol{\chi} (M,M^{'})  \cdot \boldsymbol{\mathcal{E}}(M^{'}) \right) \; .
\end{equation}
The r.h.s. of Eq.~\eqref{LRT} has been formulated in  a compact, albeit convenient notation that will
be adopted henceforth,  where
the symbol $\circ$ means both a tensorial contraction (denoted by the dot '' $\cdot $ '')
and a spacial convolution over the whole sphere. 

From standard linear response theory the  tensorial susceptibility  $ \boldsymbol{\chi} $ is then given by
\begin{equation}
\label{chi}
\boldsymbol{\chi} (M_1,M_2) = 2 \pi \beta <\widehat{\mathbf{P}} (M_1) \widehat{\mathbf{P}} (M_2)> \; ,
\end{equation}
where the thermal average $< \ldots >$ in the r.h.s. of~\eqref{chi} is now evaluated in the absence of the external 
field, $\mathcal{E}\equiv 0$. The susceptibility tensor 
$\boldsymbol{\chi} (M_1,M_2)$ can be expressed in terms of the one- and two- point correlation
functions~\eqref{correlations} as
\begin{equation}
\label{chi_bis}
\boldsymbol{\chi} (M_1,M_2)  = \boldsymbol{\chi}_S (M_1,M_2) + \boldsymbol{\chi}_D (M_1,M_2) \; ,
\end{equation}
where the ``self'' part reads
\begin{equation}
\label{chi_S}
 \boldsymbol{\chi}_S (M_1,M_2) =y \, \delta(M_1,M_2)\,  \mathbf{U}_{\mathcal{S}_2}(\mathbf{z_1}) \; ,
\end{equation}
 and
\begin{equation}
\label{chi_D}
 \boldsymbol{\chi}_D (M_1,M_2) = 2 y \rho \int_0^{2 \pi} \dfrac{d \beta_1}{2 \pi} \; \int_0^{2 \pi} \dfrac{d \beta_2}{2 \pi}
    h(1,2) \mathbf{s}_1 \mathbf{s}_2   \, .
\end{equation}
After making use of the expansion~\eqref{rotas} of $h(1,2)$ in rotational invariants one finds finally
\begin{align}
\label{chi_final}
 \boldsymbol{\chi} (M_1,M_2) & = y  \delta\left(M_1,M_2\right)\,  \mathbf{U}_{\mathcal{S}_2}\left(\mathbf{z_1}\right)   \nonumber \\
& + \dfrac{y \rho}{2} \left\{ \left[   (1- \cos \psi_{12}) h^{\Delta}(r_{12}) 
                                            +  (1+ \cos \psi_{12}) h^{D}(r_{12}) \right] \mathbf{t}_{12}(1)  \mathbf{t}_{12}(2)     \right. \nonumber \\
     & + \left. \left(  h^{\Delta}(r_{12}) -    h^{D}(r_{12})        \right) \mathbf{U}_{\mathcal{S}_2}(\mathbf{z_1})
 \cdot  \mathbf{U}_{\mathcal{S}_2}(\mathbf{z_2})\right\} \; .
\end{align}
We point out that in the limit $r_{12}$ fixed, $R \to \infty$ the above expressions reduces to that derived
in Appendix~\ref{Appendix_B} for the $2D$ euclidian space (\textit{cf} Eq.~\eqref{chi_E2_final}).

The dielectric properties of the fluid are characterized by the dielectric tensor $\boldsymbol{\epsilon}$
which enters the constitutive  relation
\begin{equation}
\label{Max}
  2 \pi \mathbf{P}= (\boldsymbol{\epsilon} - \mathbf{I}) \circ \mathbf{E} \; ,
\end{equation}
where $ \mathbf{E}$ denotes the Maxwell field and $\mathbf{I}(M_1,M_2) \equiv 
\mathbf{U}_{\mathcal{S}_2}(\mathbf{z}_1) \delta(M_1,M_2)$.
The Maxwell field $\mathbf{E}(M)$ is the sum of the  external field 
$\boldsymbol{\mathcal{E}}(M) $ and the electric field created  by the macroscopic polarization
of the fluid. 
Therefore one has 
\begin{equation}
\label{zon}
 \mathbf{E} =  \boldsymbol{\mathcal{E}} + 2 \pi \mathbf{G}_0 ^{\mathrm{mono}}\circ \mathbf{P} \; .
\end{equation}

It is generally assumed that $\boldsymbol{\epsilon}$ is a local function,
\textit{i.e.}  $\boldsymbol{\epsilon}= \epsilon \mathbf{I}$. 
More precisely, it is plausible -and we shall take it for granted- that $\boldsymbol{\epsilon}(M_1,M_2)$
is a short range function of the distance between the two points $(M_1,M_2)$, at least for a homogeneous liquid,
and one then  defines
\begin{equation}
 \epsilon \mathbf{U}_{\mathcal{S}_2}(\mathbf{z_1})=
 \int_{\mathcal{S}_2(O,R)}  d  S_2  \;  \boldsymbol{\epsilon}(M_1M_2) \; .
\end{equation}

In general
$(\boldsymbol{\epsilon}- \mathbf{I}) \neq \boldsymbol{\chi} $ since the Maxwell field  $ \mathbf{E}(M)$
and the external field  $\boldsymbol{\mathcal{E}}(M)$
do not coincide. The relation between the two fields is easily obtained from~\eqref{zon}
and usually recast as~\cite{Fulton_I,Fulton_II,Fulton_III,Caillol_4}
\begin{equation}
\label{Ee}
  \mathbf{E} =   \boldsymbol{\mathcal{E}} + \mathbf{G}^{\mathrm{mono}} \circ \boldsymbol{\sigma}  \circ  
 \boldsymbol{\mathcal{E}} \; ,
\end{equation}
where $\boldsymbol{\sigma} \equiv \boldsymbol{\epsilon} - \mathbf{I} $  and $ \mathbf{G}^{\mathrm{mono}}(M_1,M_2)$
is the macroscopic dielectric Green's function defined by the identity
\begin{equation}
\label{G}
 \mathbf{G}^{\mathrm{mono}} =  \mathbf{G}_0^{\mathrm{mono}} \circ \left(\mathbf{I}- \boldsymbol{\sigma} \circ
\mathbf{G}_0^{\mathrm{mono}} \right)^{-1} \; ,
\end{equation}
where  the inverse must be understood in the sense of operators.
It is easy to show that the electric field created at a point $M_1 \in \mathcal{S}_2(O,R)$ by a point mono-dipole $\boldsymbol{\mu}_2$
located at $M_2$
\textit{ in the presence of the dielectric medium } is then given by $2 \pi \mathbf{G}^{\mathrm{mono}}(M_1,M_2) \cdot \boldsymbol{\mu}_2$.
This remark enlightens the physical meaning of the macroscopic Green's function.

Combining Eqs.~\eqref{LRT}, ~\eqref{Max}, and ~\eqref{Ee} yields Fulton's relation
\begin{equation}
 \label{F_rel}
\boldsymbol{\chi} = \boldsymbol{\sigma}  +  \boldsymbol{\sigma} \circ  \mathbf{G}^{\mathrm{mono}} \circ \boldsymbol{\sigma} \; .
\end{equation}
To go further one has to compute  the macroscopic Green's function $ \mathbf{G}^{\mathrm{mono}}$.
Our starting point
is the following identity, proved in appendix~\ref{Appendix_A} :
\begin{equation}
 \mathbf{G}_0^{\mathrm{mono}} \circ  \mathbf{G}_0^{\mathrm{mono}}  = - \mathbf{G}_0^{\mathrm{mono}}  \; .
\end{equation}
Therefore $ - \mathbf{G}_0^{\mathrm{mono}}$ is a projector and has no inverse. Assuming the locality of $\boldsymbol{\sigma}$
one is then led to search the inverse  $\left(\mathbf{I}- \boldsymbol{\sigma} \circ
\mathbf{G}_0^{\mathrm{mono}} \right)^{-1}$ in the r.h.s. of~\eqref{G} under the form $a \mathbf{I} + b \mathbf{G}_0^{\mathrm{mono}} $
where $a$ and $b$ are numbers (or local operators). By identification one finds $a=1$ and $b=\sigma/(1+ \sigma)$
yielding for  $ \mathbf{G^{\mathrm{mono}}}$ the simple (and expected) expression
\begin{equation}
  \mathbf{G}^{\mathrm{mono}}= \mathbf{G}_0 ^{\mathrm{mono}}/(1+ \sigma) \equiv \mathbf{G}_0^{\mathrm{mono}}/\epsilon \; .
\end{equation}

The results derived above allow us  to recast Fulton's relation~\eqref{F_rel} under its final form
\begin{equation}
 \label{F_rel_bis} 
(\epsilon-1)  \mathbf{I}(M_1,M_2) + \frac{(\epsilon-1)^2}{\epsilon} \;  \mathbf{G}_0^{\mathrm{mono}} (M_1,M_2)
 = \boldsymbol{\chi} (M_1,M_2)  \; .
\end{equation}
We stress that the above equation has been obtained under the assumption of the locality
of the dielectric tensor  $\boldsymbol{\epsilon}(M_1,M_2)$. Therefore it should be valid only asymptotically,
\textit{i.e.} for points $(M_1,M_2)$ at a mutual distance $r_{12}$ larger then the range 
$\xi$ of $\boldsymbol{\epsilon}(M_1,M_2)$.

Expressions for the dielectric constant  are  obtained
from Eq.~\eqref{F_rel_bis} by space integration. Following Refs.~\cite{Berend,Caillol_4,Tr_Cai} one integrates
both sides of Eq.~\eqref{F_rel_bis} and then takes the trace. The integration of $M_2$ is performed
over a cone of axis $\mathbf{z}_1$ and aperture $\psi_0$ and then $M_1$ is
integrated over  the whole sphere $\mathcal{S}_2(O,R)$. The singularity
of the dipolar Green's function  $\mathbf{G}_0(M_1,M_2)$ for $\psi_{12} \sim 0$ must be carefully taken into
account and this delicate point is detailed in appendix~\ref{Appendix_A} (see Eq.~\eqref{int1}). One finds finally
  \begin{equation}
\label{cloc_mono}
   \frac{\epsilon -1}{\epsilon} +  \frac{(\epsilon -1)^2}{4 \epsilon} (1+\cos \psi_0) = \mathbf{m}^2(\psi_0) 
 \;   \text{  with  }  0< \psi_0<\pi \; ,
  \end{equation}
where the dipolar fluctuation $ \mathbf{m}^2(\psi_0)$ reads as
  \begin{equation}
\label{epsi_mono}
   \mathbf{m}^2(\psi_0) = \frac{ \pi \beta \mu^2}{S}  < \sum_i^N \sum_j^N \mathbf{s}_i \cdot  \mathbf{s}_j \, \Theta (\psi_0 -\psi_{ij})> \; ,
  \end{equation}
where $\Theta(x)$ is the Heaviside step-function ($\Theta(x)=0$ for $x<0$ and  $\Theta(x)=1$ for $x>0$).

We  have thus obtained a family of formula depending on parameter $0< \psi_0 < \pi$; clearly they should be valid  only if $R\psi_0$ is large
when compared to the range $\xi$ of the dielectric constant. The numerical results of Sec.~\eqref{Simulations}
show that this range is of the order of a few atomic diameters.
It is also important to note that for $\psi_0=\pi$ Eq.~\eqref{cloc_mono} involves the fluctuations
of the total $3 \mathrm{D}$ dipole moment of the system.  However, the resulting formula \textit{i.e.}
$(\epsilon -1)/\epsilon=\mathbf{m}^2(\pi)=\pi \beta<\mathbf{M}^2>/S $ (with 
$\mathbf{M} = \sum_i \boldsymbol{\mu}_i$ the total $3D$
dipole moment of the sphere), albeit simple, 
is not adapted for numerical applications since, for  large values  of the dielectric constant,
a reasonable numerical error on $\epsilon$ requires a determination of
$\mathbf{m}^2(\pi) $ with an impractical precision. 
The choice $\psi_0=\pi/2$ yields the less simple formula
$(\epsilon-1) (\epsilon+3)/ \epsilon =4 \, \mathbf{m}^2(\pi/2) $ which however allows, by contrast, a precise
determination of $\epsilon$.  Indeed, let $\delta \epsilon$ be the error on $\epsilon$, then, for  high values of the dielectric constant 
the errors on $\mathbf{m}^2(\pi/2)$ and  $ \epsilon$ are roughly   proportional, \textit{i.e.}
 $\delta \epsilon\sim  4\,  \delta \mathbf{m}^2(\pi/2)$.

The fluctuation $\mathbf{m}^2(\psi_0) $ is  related to  the so-called  Kirkwood factor $G^K(\psi_0)$.
according to the relation $\mathbf{m}^2(\psi_0) =  y G^K(\psi_0)$ with
 \begin{equation}
\label{Kirk}
  G^K(r) = 1 + \rho \pi \, R^2 \int_0^{\psi_0}d \psi  \;  \sin \psi \,  h^{\mathrm{Kirk.}}(R \psi) \; ,
 \end{equation}
where 
 \begin{equation}
   h^{\mathrm{Kirk.}}(r=R \psi) =\frac{1}{2}\, (1 + \cos \psi ) h^{\Delta}(r) -  \frac{1}{2}\, (1 - \cos \psi ) h^{D}(r) \; .
 \end{equation}

Fulton's relation~\eqref{F_rel_bis} can also be used to determine the asymptotic
behavior of the projections   $h^{\Delta}(r)$ and   $h^{D}(r)$ the pair-correlation function $g(1,2)$.
By comparing Eqs.~\eqref{chi_final} and~\eqref{F_rel_bis} one obtains readily that, \textit{asymptotically},
\textit{i.e.}  for large $r=R \psi$ and $\psi < \pi$,  one has
\begin{subequations}
\label{assym_mono}
  \begin{align}
   h^{\Delta}_{\text{asymp.}}(r) &  \sim 0   \; , \\
   h^{D}_{\text{asymp.}}(r)          & \sim \frac{(\epsilon-1)^2}{y \rho \epsilon} \frac{1}{2 \pi R^2 }\frac{1}{1  -  \cos \psi} \; ,  \\
  h^{\mathrm{Kirk.}}_{\text{asymp.}}(r ) &  \sim - \frac{(\epsilon-1)^2}{y \rho \epsilon} \frac{1}{4 \pi R^2} 
 \end{align}
 \end{subequations}
We note that  $h^{\Delta}(r)$ has no tail and is thus a short-range function. It thus presents the same behavior as
in the $2D$ Euclidean space (see Refs.\cite{Orsay_1,these} and the new analysis in appendix~\ref{Appendix_B}
also based on Fulton's relation). We also note that, in the thermodynamic
 limit $R \to \infty$ \textit{and} with $r \gg \xi$ fixed but large, one recovers the expected
Euclidian behavior $ h^{D}_{\text{asymp.}}(r) \sim (\epsilon-1)^2/( \pi y \rho \epsilon) \times 1/r^2$ valid for
an infinite system without boundaries at infinity( see \textit{e.g.}~\cite{Orsay_1,these} and Eq.~\eqref{asy} of
appendix~\ref{Appendix_B}).
The Kirkwood's function $ h^{\mathrm{Kirk.}}(r)$ exhibits a  constant tail which tends to zero as
the inverse of the surface $S = 4 \pi R^2$ of the system. However the integration of this tail over the volume
gives a non-zero contribution to the dielectric constant. In the limit  $r = R \psi \gg \xi$ one has  
 \begin{align}
  G^K(\psi_0) &=   G^K_{\infty} + \pi \rho R^2 \int_0^{\psi_0} d\psi \sin \psi  \,  h^{\mathrm{Kirk.}}_{\text{asymp.}}(R \psi ) \nonumber \\
                    &=  G^K_{\infty}  - \dfrac{(\epsilon -1)^4}{4 y \epsilon} \, (1-\cos \psi_0) \; ,
 \end{align}
where $G^K_{\infty}$ is the Euclidian Kirkwood factor
\begin{equation}
 G^K_{\infty} =1 + \frac{\rho}{2} \, \int_0^{\infty}dr \,  2 \pi r \,  h^{\Delta}_{\infty}(r )  \; ,
\end{equation}
where we have noted that, in the thermodynamic limit, $r$ fixed and $R \to \infty$, $ h^{\mathrm{Kirk.}}(r)
=h^{\Delta}_{\infty}(r)$. Reporting this asymptotic expression of the Kirkwood factor in 
Eq.~\eqref{epsi_mono} one recovers
the formula valid for the Euclidian plane which reads as
 \begin{equation}
 \label{epsilon_2}
\dfrac{(\epsilon -1)(\epsilon+1)}{2 \epsilon}=y  G^K_{\infty} \; .
\end{equation} 
which coincides with the Eq.~\eqref{epsilon} obtained in the appendix~\ref{Appendix_B}
for the infinite  plane $E_2$;  note that  Eq.~\ref{epsilon_2} can be derived by a whole host of 
other methods, cf Refs~\cite{Orsay_1,these}.
\subsubsection{Bi-dipoles}
\label{Fult_bi}
The passage from mono- to bi-dipoles requires some adjustments. First, the appropriate bare
Green's function is now $\mathbf{G}_0^{\mathrm{bi}}$ as defined at Eqs.~\eqref{G0_bi}. Secondly
the operation $\circ $ defined at Eq.~\eqref{LRT} now involves a spacial integration the support of which
is only the northern hemisphere $\mathcal{S}^{+}_2(O,R)$ rather than the entire
surface of the sphere. For instance,
as shown in appendix~\ref{Appendix_A} one has 
\begin{align}
\left[ \mathbf{G}_0^{\mathrm{bi}} \circ  \mathbf{G}_0^{\mathrm{bi}}\right](M_1,M_2)  &=  \int_{\mathcal{S}^{+}_2(O,R)} 
dS \, \mathbf{G}_0^{\mathrm{bi}}(M_1,M) \cdot  \mathbf{G}_0^{\mathrm{bi}}(M,M_2)     \nonumber \\
                                                                                                  &= -\mathbf{G}_0^{\mathrm{bi}}(M_1,M_2)
\end{align}
for points points $M_1$ and $M_2$ belonging to the northern hemisphere.
Therefore, reasoning as in Sec.~\ref{Fult_mono}, we find for the dressed Green's function, asymptotically
 \begin{equation}
  \label{G_bi}
 \mathbf{G}^{\mathrm{bi}} =  \mathbf{G}_0^{\mathrm{bi}} \circ \left(\mathbf{I}- \boldsymbol{\sigma} \circ
\mathbf{G}_0^{\mathrm{bi}} \right)^{-1} \;  = \mathbf{G}_0^{\mathrm{bi}}/\epsilon \; .
 \end{equation}
Clearly the susceptibility tensor     $\boldsymbol{\chi}_D (M_1,M_2) $   keeps the same expressions~\eqref{chi}
and~\eqref{chi_final} with the restriction that the two points $M_1$ and $M_2$ both belong
to the northern hemisphere $\mathcal{S}^{+}_2(O,R)$. As in Sec.~\ref{Fult_mono} the dielectric
constant  can be  obtained from Fulton's
relation $\boldsymbol{\chi} = \boldsymbol{\sigma}  
+  \boldsymbol{\sigma} \circ  \mathbf{G}^{\mathrm{bi}} \circ \boldsymbol{\sigma}$ and reads now
  \begin{equation}
\label{cloc_bi}
   \frac{\epsilon -1}{\epsilon} +  \frac{(\epsilon -1)^2}{2 \epsilon} \, \cos \psi_0 = \mathbf{m}^2(\psi_0) 
    \; 
 \text{  with  }   0<  \psi_0 <\pi/2 \; ,
  \end{equation}
where the fluctuation $\mathbf{m}^2(\psi_0)$ is still given by
  \begin{equation}
\label{eps_bi}
   \mathbf{m}^2(\psi_0) = \frac{ \pi \beta \mu^2}{S}  < \sum_i^N \sum_j^N \mathbf{s}_i \cdot  \mathbf{s}_j \, \Theta (\psi_0 -\psi_{ij})> \; ,
  \end{equation}
but with $S=2 \pi R^2$ (surface of the northern hemisphere) and $0 < \psi_0 <\pi/2$. The delicate mathematical
integration of the Green's function   $ \mathbf{G}^{\mathrm{bi}}$ required to derive Eq.~\eqref{eps_bi} is explained in appendix~\ref{Appendix_A}.

We can note that for $\psi_0 =\pi/2$, \textit{i.e.} when the fluctuation of the total
dielectric moment of the sample is taken into account one still has $(\epsilon -1)/\epsilon=\pi \beta <\mathbf{M}^2>/S$
with $\mathbf{M}= \sum_i \boldsymbol{\mu}_i$ the total $3D$ dipole moment of the northern hemisphere.
Quite remarkably one obtains for $\psi_0=\pi/3$ the relation
$(\epsilon-1) (\epsilon+3)/ \epsilon = 4 \, \mathbf{m}^2(\pi/3) $ which is formally identical to that obtained
for mono-dipoles with the choice  $\psi_0=\pi/2$. It is not a surprise since, in both cases, $\mathbf{m}^2(\psi_0)$ accounts
for the dipole fluctuation of half the total domain available to the dipoles of the system.

The asymptotic behavior of the pair correlation function is obtained in the same vein as in Sec.~\ref{Fult_mono},
\textit{i.e.}  for large $r=R \psi$ and $\psi < \pi/2$,  one has
\begin{subequations}
\label{assym_bi}
  \begin{align}
   h^{\Delta}_{\text{asymp.}}(r) &  \sim  -\frac{(\epsilon-1)^2}{y \rho \epsilon} \frac{1}{2 \pi R^2 }  \frac{1}{1  +  \cos \psi}   \; , \\
   h^{D}_{\text{asymp.}}(r)          & \sim \frac{(\epsilon-1)^2}{y \rho \epsilon} \frac{1}{2 \pi R^2 }\frac{1}{1  -  \cos \psi} \; ,  \\
  h^{\mathrm{Kirk.}}_{\text{asymp.}}(r ) &  \sim - \frac{(\epsilon-1)^2}{y \rho \epsilon} \frac{1}{2 \pi R^2} 
 \end{align}
 \end{subequations}

In the  limit $R \to \infty$ \textit{and} with $r \gg \xi$ fixed but large, one recovers once again the expected
Euclidian behavior $ h^{D}_{\text{asymp.}}(r) \sim (\epsilon-1)^2/( \pi y \rho \epsilon) \times 1/r^2$.
By contrast, in the same limit, one obtains  that $ h^{\Delta}_{\text{asymp.}}(r)\sim  - (\epsilon-1)^2/(4 \pi y \rho \epsilon) \times 1/R^2$
which tends to zero for the infinite system for which $R \to \infty$. This behavior is in agreement with the expected short range behavior of 
the projection $ h^{\Delta}(r)$ in the $2D$ infinite Euclidian plane (cf Refs.~\cite{Orsay_1,these} and the appendix~\ref{Appendix_B}).
The Kirkwood's function   $h^{\mathrm{Kirk.}}_{\text{asymp.}}(r )$ presents the same constant tail $- (\epsilon-1)^2/(y \rho \epsilon) \times 1/S$
as in the mono-dipole cases. Finally, in the thermodynamic limit,  the Kirkwood's factor behaves as
$G^K(\psi_0) =   \mathbf{m}^2(\psi_0)/ y   =   G^{K}_{\infty} - (\epsilon-1)^2 \,(1-\cos \psi_0)/(2 y \epsilon)$ ;
by inserting this expression in Eq.~\eqref{cloc_bi}
one recovers, as for mono-dipoles,  the formula~\eqref{epsilon_2} valid for the  Euclidian plane $E_2$.

\section{Simulations}
\label{Simulations}
The only published MC simulations of the $2D$ DHS fluid the author is aware of are those
of Morriss and Perram~\cite{Morris} published 30 years ago.
Their simulation cell is a square with periodic boundary conditions and the authors
make a correct  use of Ewald dipolar potentials (an alternative
version of that used in their paper, however non implemented
in actual numerical simulations, is discussed in Ref.~\cite{Perram}).
We have retained one of their points, \textit{i.e.} our state~I : $( \rho^{*} =0.7$, $\mu^{*2} = 2 )$
as a benchmark. We also report MC data for a second reference state, 
our state~II $ \equiv ( \rho^{*} =0.6$, $\mu^{*2} = 4 )$ characterized by a much higher
dielectric constant. Both states undoubtly belongs to  a stable fluid phase as it will be discussed below.

We performed standard MC simulations of a DHS fluid in the canonical ensemble
with single particle displacement moves (translation and rotation). Each new configuration is thus generated by
the trial  displacement of \emph{a randomly chosen single} dipole by means of  a new
algorithm explained in appendix~\ref{Appendix_C}.

We considered both mono- and bi-dipoles on the sphere and studied finite size
effects on the thermodynamic, structural and dielectric properties of the fluid. 
We studied systems of 
$N=250, 500, 1000, 2000,  \text{ and } 4000$ particles and typically we generated
$N_{\mathrm{Nconf}}=16 \, 10^9$ configurations for each considered state.

Our results are resumed in Table~\ref{I} (state~I) and Table~\ref{II} (state~II).
The errors reported in these tables correspond to two standard deviations 
in a standard statistical  block analysis of the MC data~\cite{Frenkel}. 

\begin{figure}[t!]
\centering
\begin{tabular}{ c }
\includegraphics[scale=0.50]{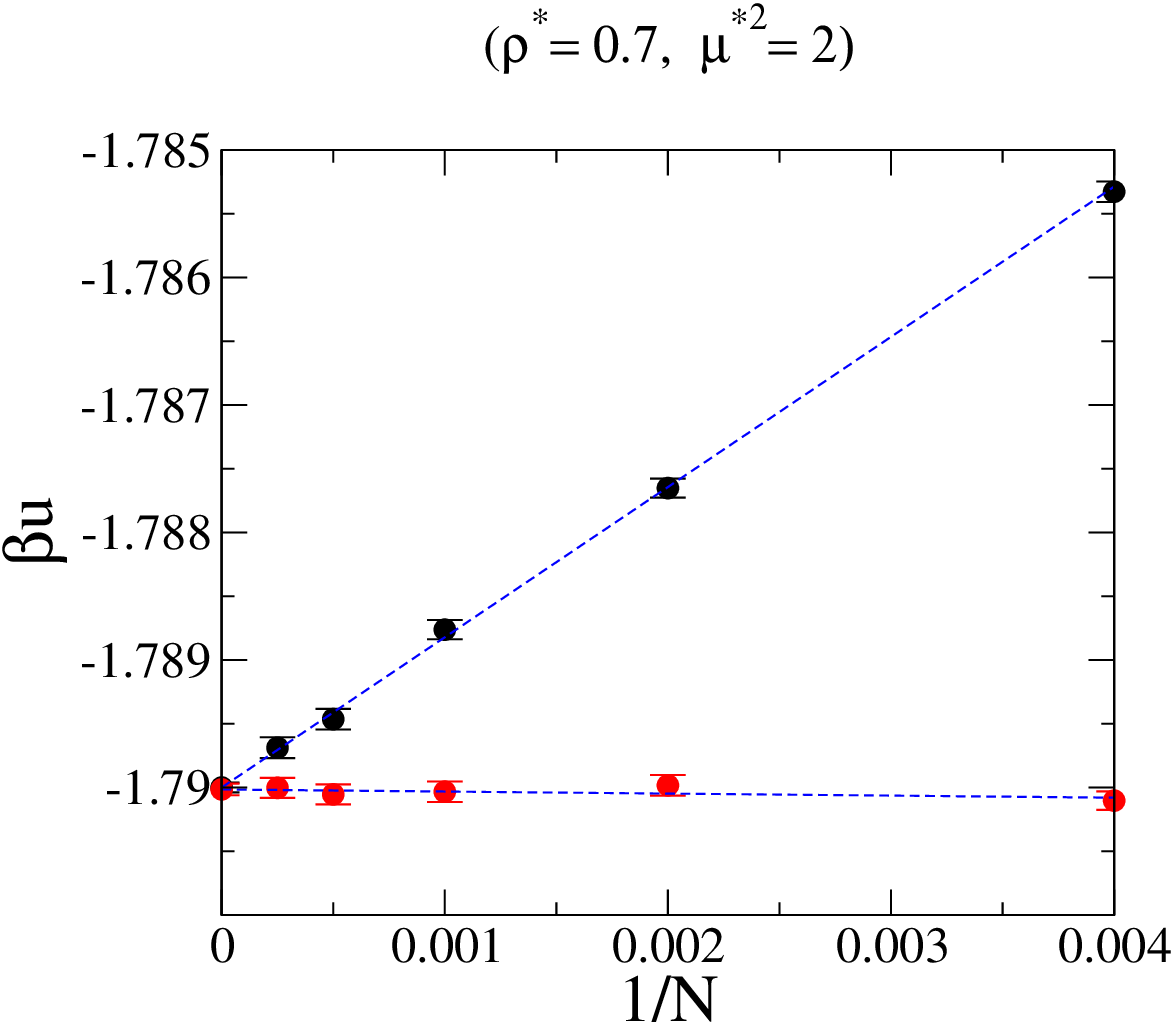}  \\
\vspace{2em} \\
\includegraphics[scale=0.50]{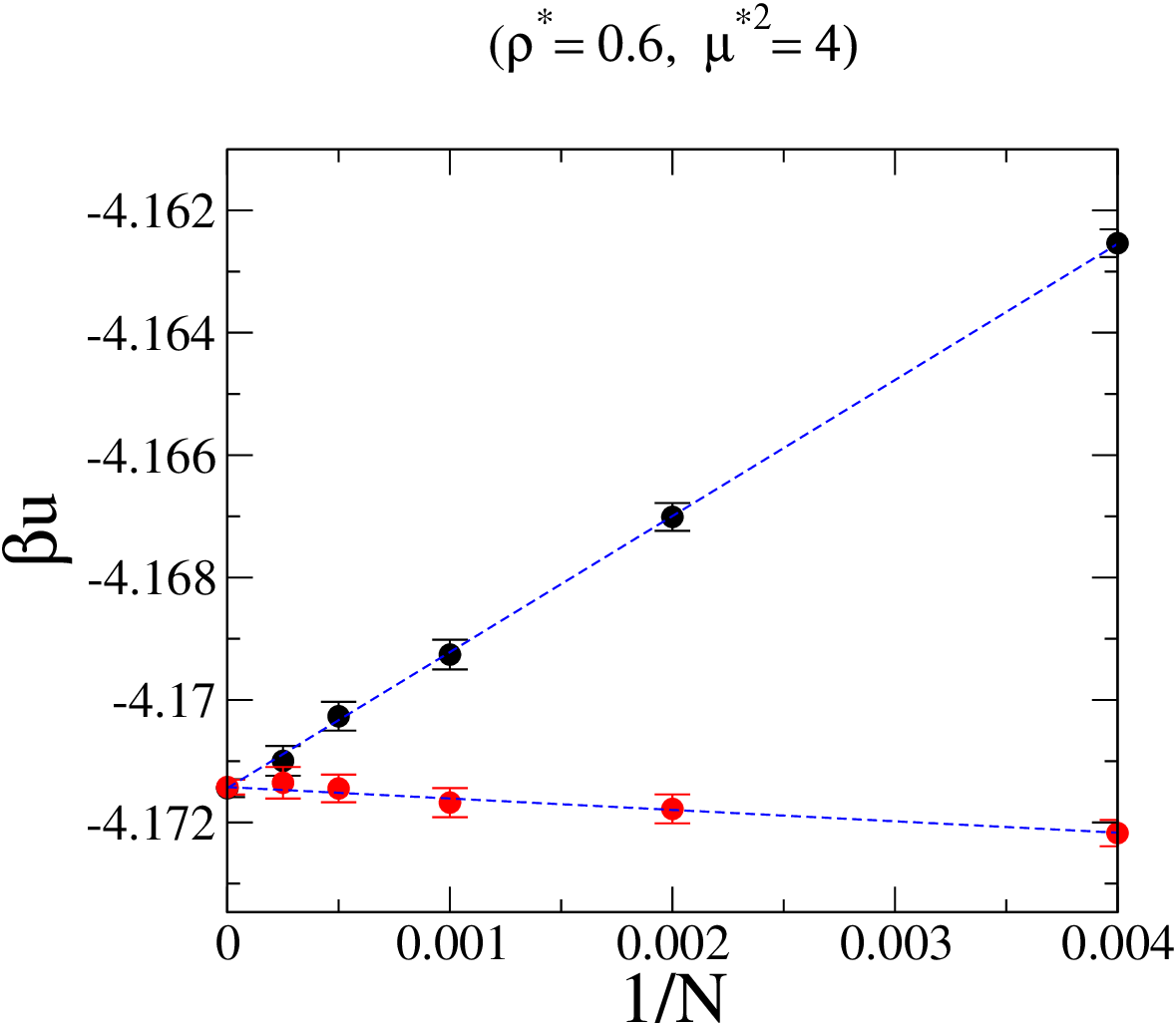}  \\
\end{tabular}
\caption{Dimensionless MC internal energy  $\beta u$ versus the inverse
number of particles $1/N$  for the
states ($\rho^{*} =0.7$, $\mu^{*2} = 2$) (top) and ($\rho^{*} =0.6$, $\mu^{*2} = 4$) (bottom).
Black circles : mono-dipoles, red circles : bi-dipoles. The error bars correspond to two
standard deviations. Blue dashed lines : linear regressions.}
\label{Fig1}
\end{figure}

As apparent in Fig.~\ref{Fig1} the reduced internal $\beta u$ energies converge linearly with $1/N$ to
their thermodynamic limit $\beta u_{\infty}$ for systems involving more than $N=250$ particles. This is true for systems
of mono- and bi-dipoles. This allows
a  determination of $\beta u_{\infty}$  with a precision of $\sim 3\, 10^{-5}$. The asymptotic values obtained
for mono- and bi-dipoles are identical within the error bars. Clearly the convergence
towards the thermodynamic limit is faster for bi-dipoles. Probably because, for a given density and
number of particles, the radius $R$ of the sphere is $\sqrt{2}$ larger for a system of bi-dipoles and therefore curvature
effects are smaller.
We note that   our finding $\beta u_{\infty} = -1.79002 \pm 5 10^{-5}  $ compares reasonably well with that obtained by Morriss and Perram
for samples of $N=100$ particles which lie in the interval $-1.795< \ldots < -1.780$ 
depending on the type of boundaries surrounding the system at infinity~\cite{Morris}.

Size effects on the short range part of the projections $g^{00}(r)$, $h^{D}(r)$, and $h^{\Delta}(r)$ are very small and cannot be appreciated
on the graphs of the functions. To give an idea of these effects we give the contact value  of these three projections 
in the tables. These values, which result from a Lagrange's polynomial interpolation, are much less precise than the values 
obtained for the energy.
The compressibility factors $Z$ were deduced from Eq.~\eqref{Z} and are also reported in the tables. The thermodynamic
limit of these quantities do not seem to obey a simple linear regression with $1/N$. However, the low precision 
on the contact value $g^{00}(\sigma)$ precludes an unambiguous study of the thermodynamic limit in this case.
We can infer from our MC data that, for state~I one has in the thermodynamic limit  $Z_{\infty}=3.98 \pm 0.01$
a value once again not far from that  of Ref.~\cite{Morris} $Z \sim4.06$ obtained for a small system
of a $N=100$ dipoles.

\begin{figure}[t!]
\centering
\begin{tabular}{ c c  c }
\includegraphics[scale=0.45]{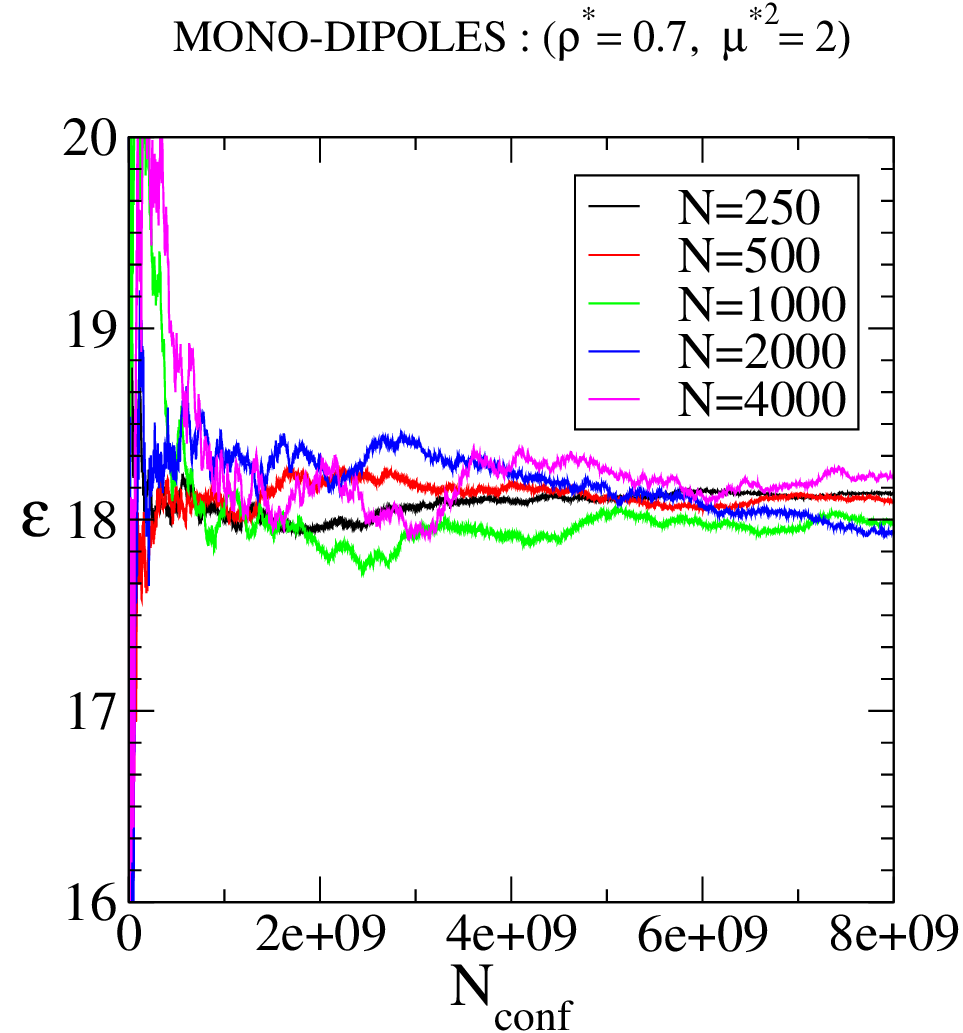} &\hspace{1.0em} & \includegraphics[scale=0.45]{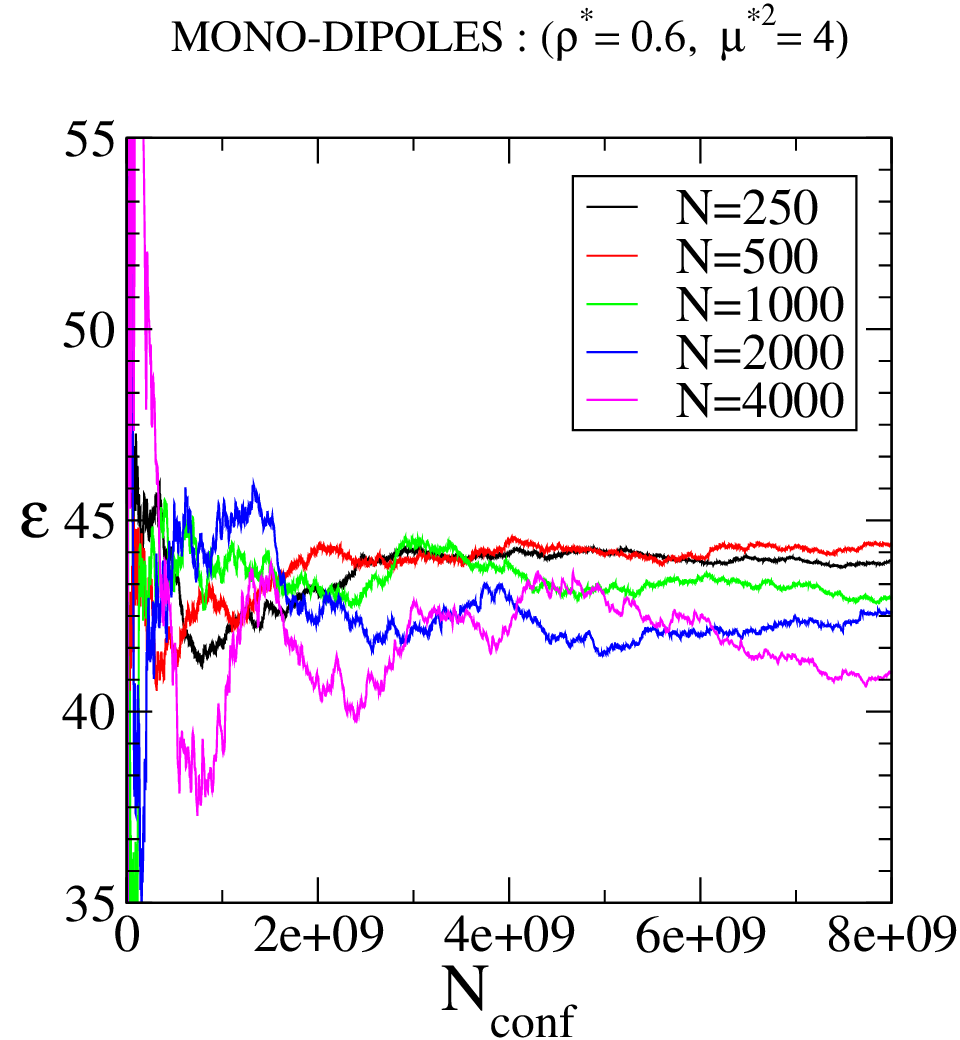} \\
\vspace{1.0em} \\
\includegraphics[scale=0.45]{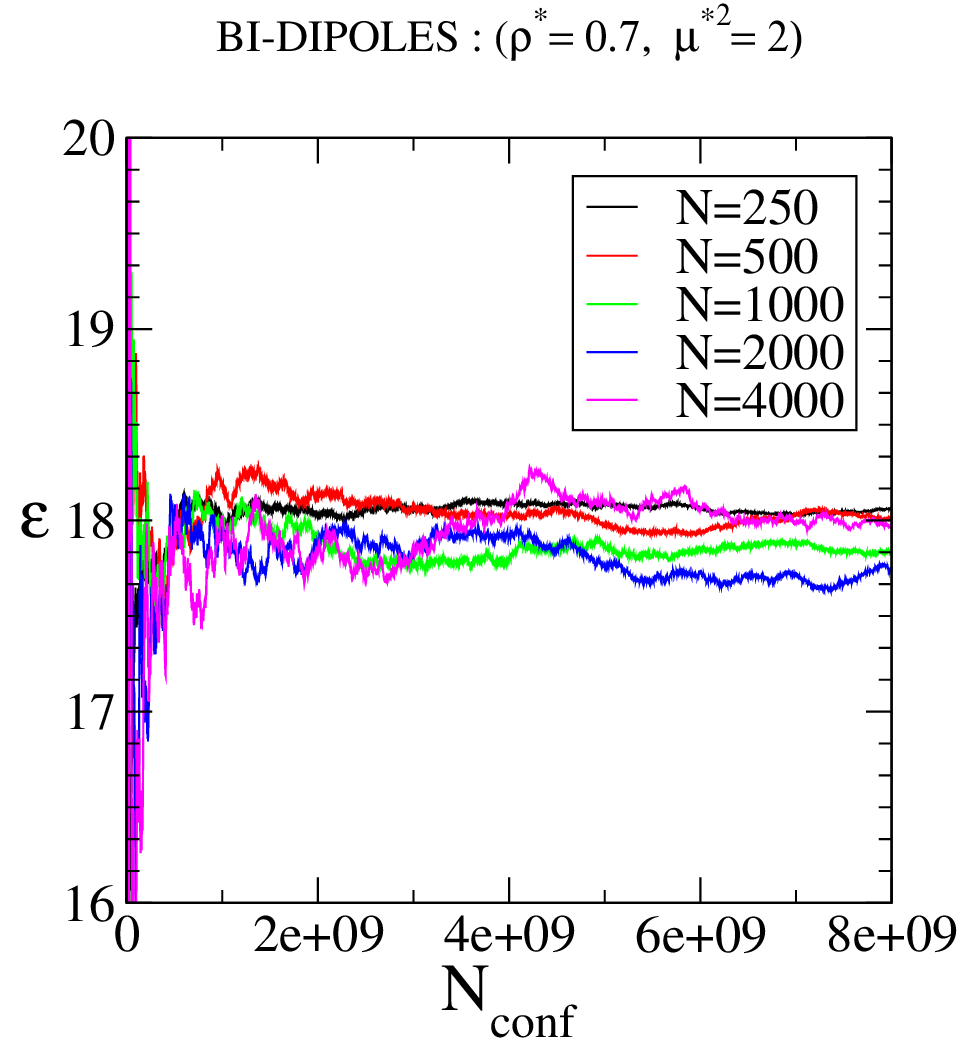} &\hspace{1.0em} & \includegraphics[scale=0.45]{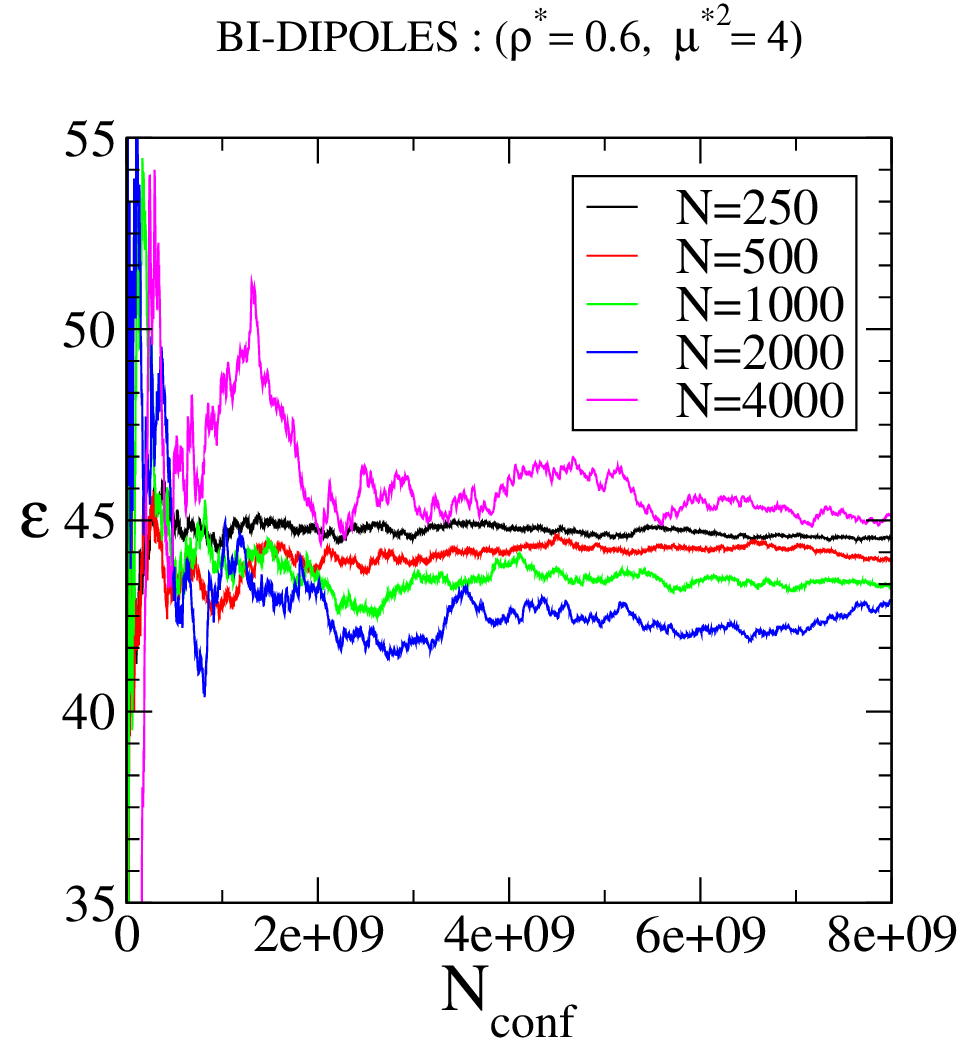}  \\
\end{tabular}
\caption{Cumulated dielectric constant $\epsilon$ for  the states  I $\equiv$ ($\rho^{*} =0.7$, $\mu^{* 2} = 2$) (left)
and  II $\equiv$($\rho^{*} =0.6$, $\mu^{* 2} = 4$) (right) as a function
of the number of configurations $N_{\mathrm{conf}}$. Top : mono-dipoles, bottom : bi-dipoles.}
\label{Fig2}
\end{figure}

The dielectric constants were obtained from Eq.~\eqref{epsi_mono} with $\psi_0=\pi/2$ for mono-dipoles
and from Eq.~\eqref{eps_bi} with $\psi_0=\pi/3$ for dipoles. As apparent on Figs.~\ref{Fig2} the numerical uncertainties 
on $\epsilon$ are large and seem even to increase
with system size but do not mask  significant finite size effects. As for the pressure, it was
impossible to deduce from our MC data a  convergence law of $\epsilon(N)$
for $N \to \infty$ despite the huge number of MC steps involved. However we
can claim that, for state~I one has $\epsilon_{\infty}=18.0 \pm 0.2$,  which differs significantly from
 the value reported by Morriss and Perram $\epsilon = 16.0 \pm 0.5$.

\begin{figure}[t!]
\centering
\begin{tabular}{ c c c  }
\includegraphics[scale=0.45]{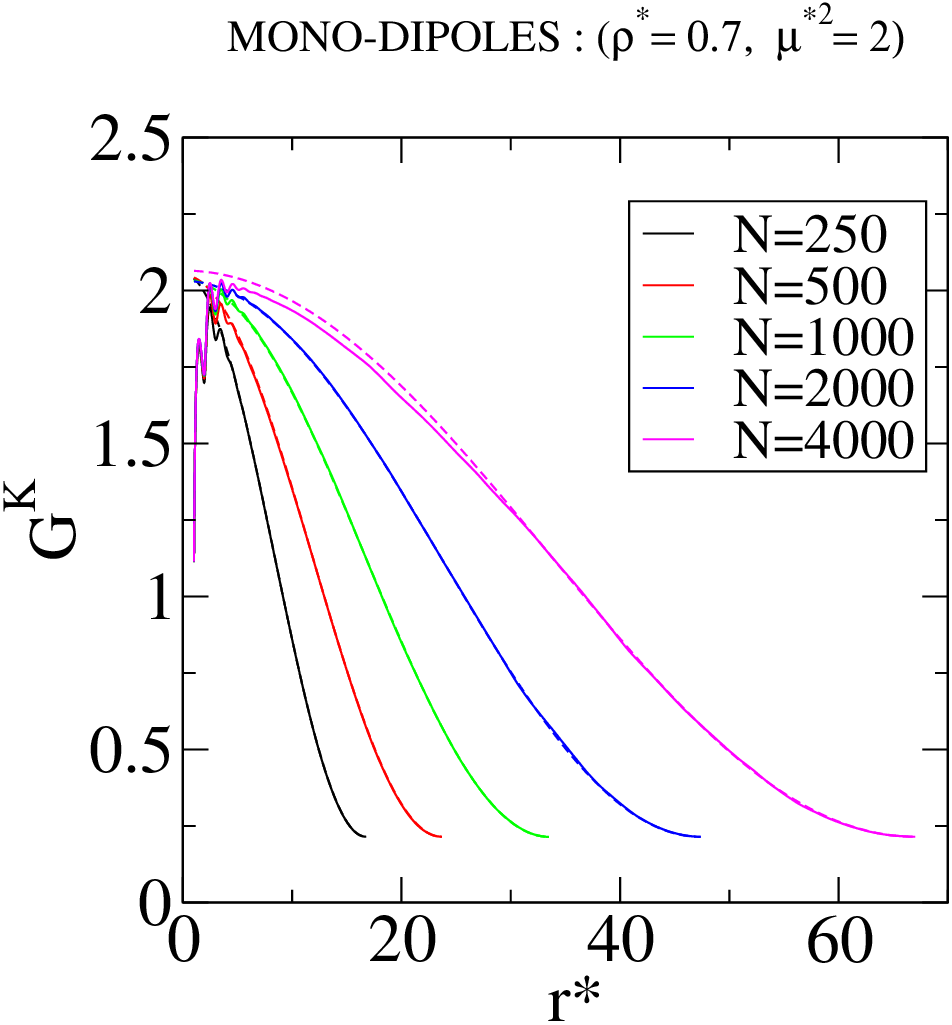} &\hspace{1.0em} & \includegraphics[scale=0.45]{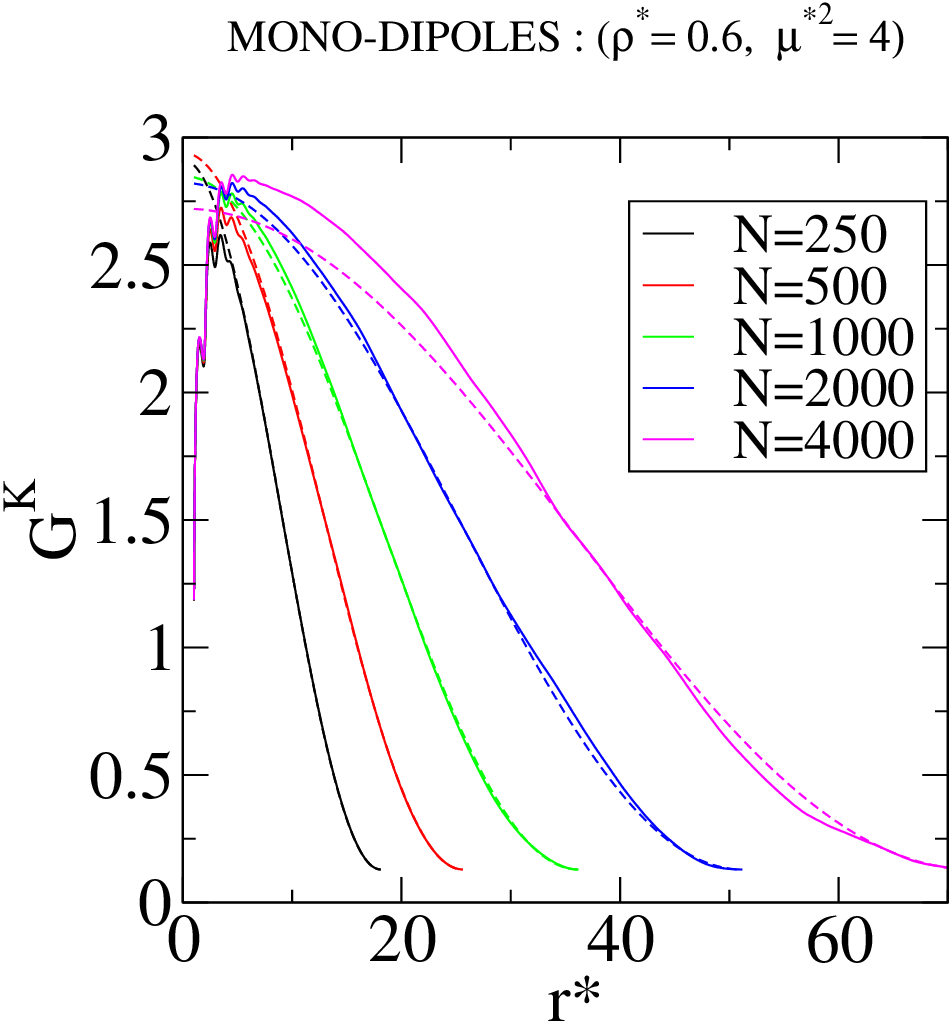} \\
\vspace{2.5em} \\
\includegraphics[scale=0.45]{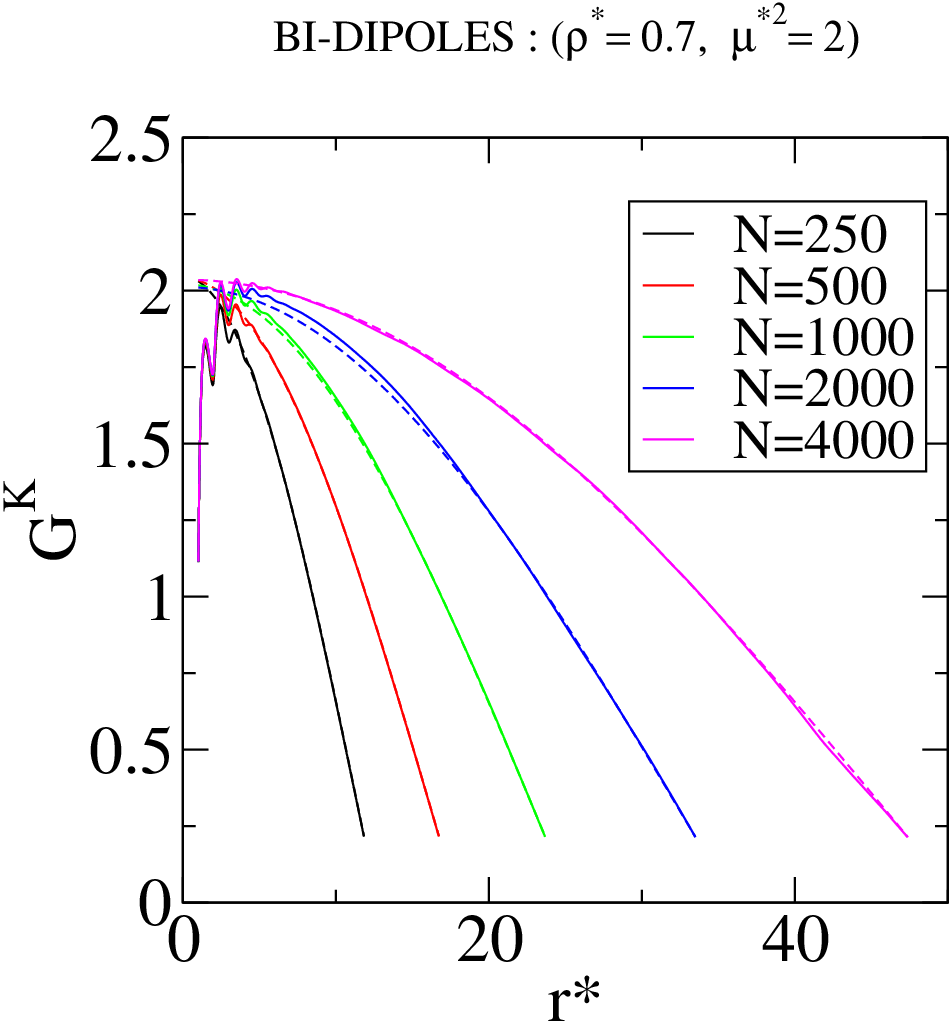} & \hspace{1.0em}  &  \includegraphics[scale=0.45]{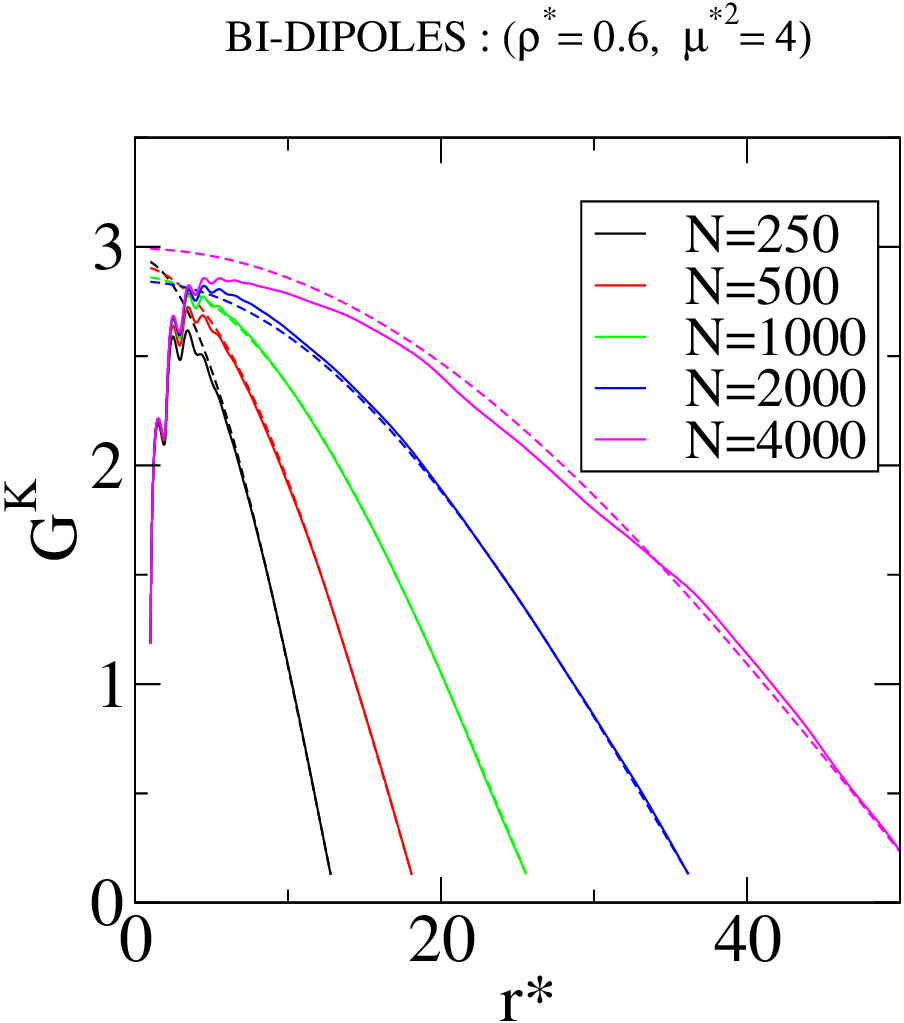} \\
\end{tabular}
\caption{Kirkwood's factor $G^{K}$ for the states  I$\equiv$($\rho^{*} =0.7$, $\mu^{* 2} = 2$) (left)
and  II$\equiv$($\rho^{*} =0.6$, $\mu^{* 2} = 4$) (right)
and various numbers $N$ of particles as a function
of the reduced distance $r^*=r/\sigma$. Top : mono-dipoles, bottom : bi-dipoles. Solid lines : MC data,
dashed lines : asymptotic behaviors given by Eq.~\eqref{cloc_mono} for mono-dipoles and Eq.~\eqref{cloc_bi} for bi-dipoles.}
\label{Fig3}
\end{figure}

In order to test the validity of Fulton's theory one can examine Fig.~\ref{Fig3} which displays
the Kirkwood factor $G^{K}(r)$ as a function of the distance $r= R\psi_0$ which defines the radius
of the cap where the partial electric moment fluctuations are taken into account. As discussed in Sec.~\ref{FULT},
for $r >\xi$ ($\xi$ range of the dielectric tensor) this function should be given by the asymptotic
behaviors given by Eq.~\eqref{cloc_mono} for mono-dipoles and Eq.~\eqref{cloc_bi} for bi-dipoles.
The dielectric constant used for these asymptotic forms are those given in Tables~\ref{I} and~\ref{II}.
The excellent agreement beetween the MC data for $G^{K}(r)$ and the theoretical
predictions even yields  an estimation for the parameter $\xi$, \text{i.e.} $\xi \sim 7 \sigma$ for both states~I and~II..

\begin{figure}[t!]
\centering
\begin{tabular}{ c }
\includegraphics[scale=0.50]{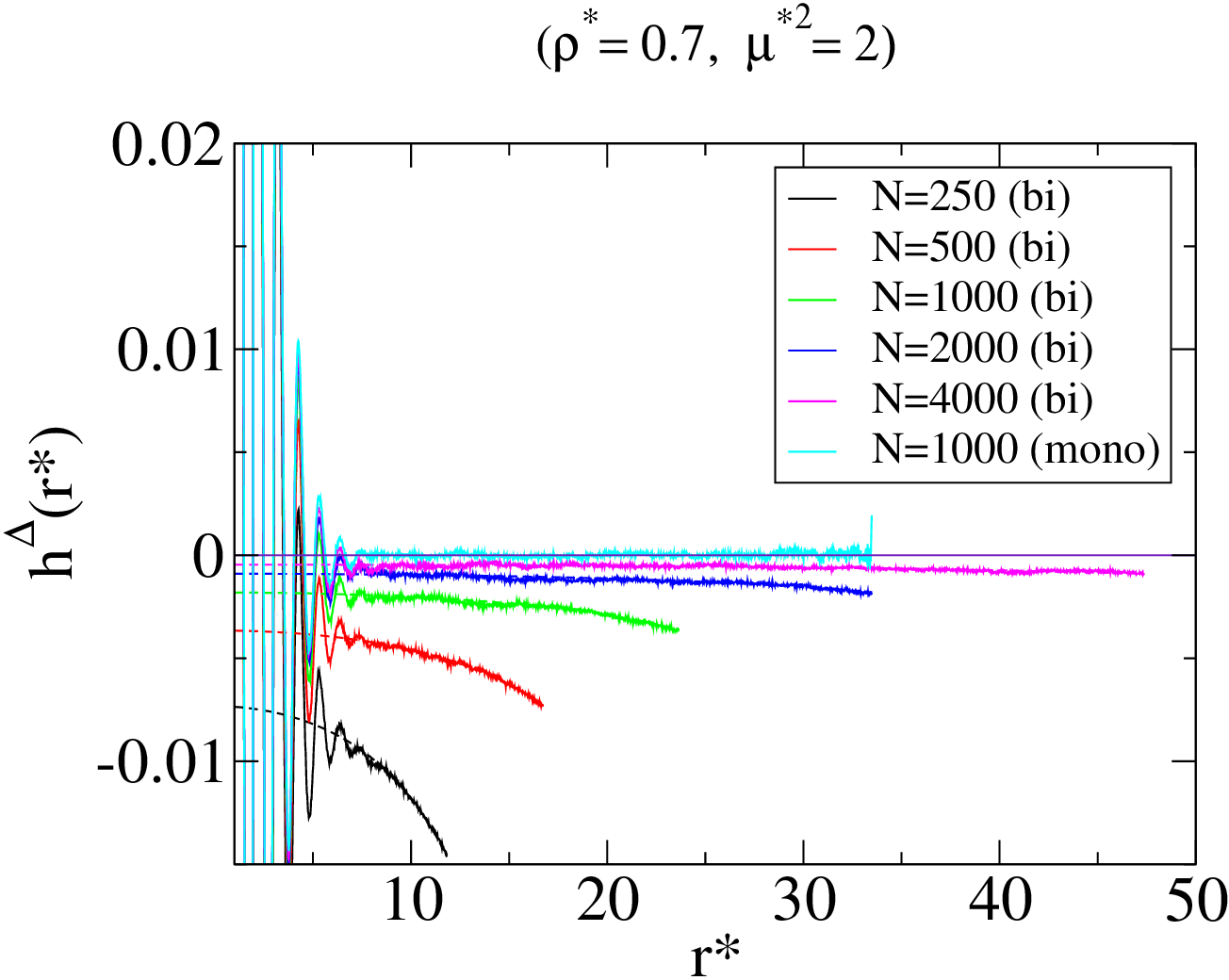}  \\
\vspace{1.0em} \\
\includegraphics[scale=0.50]{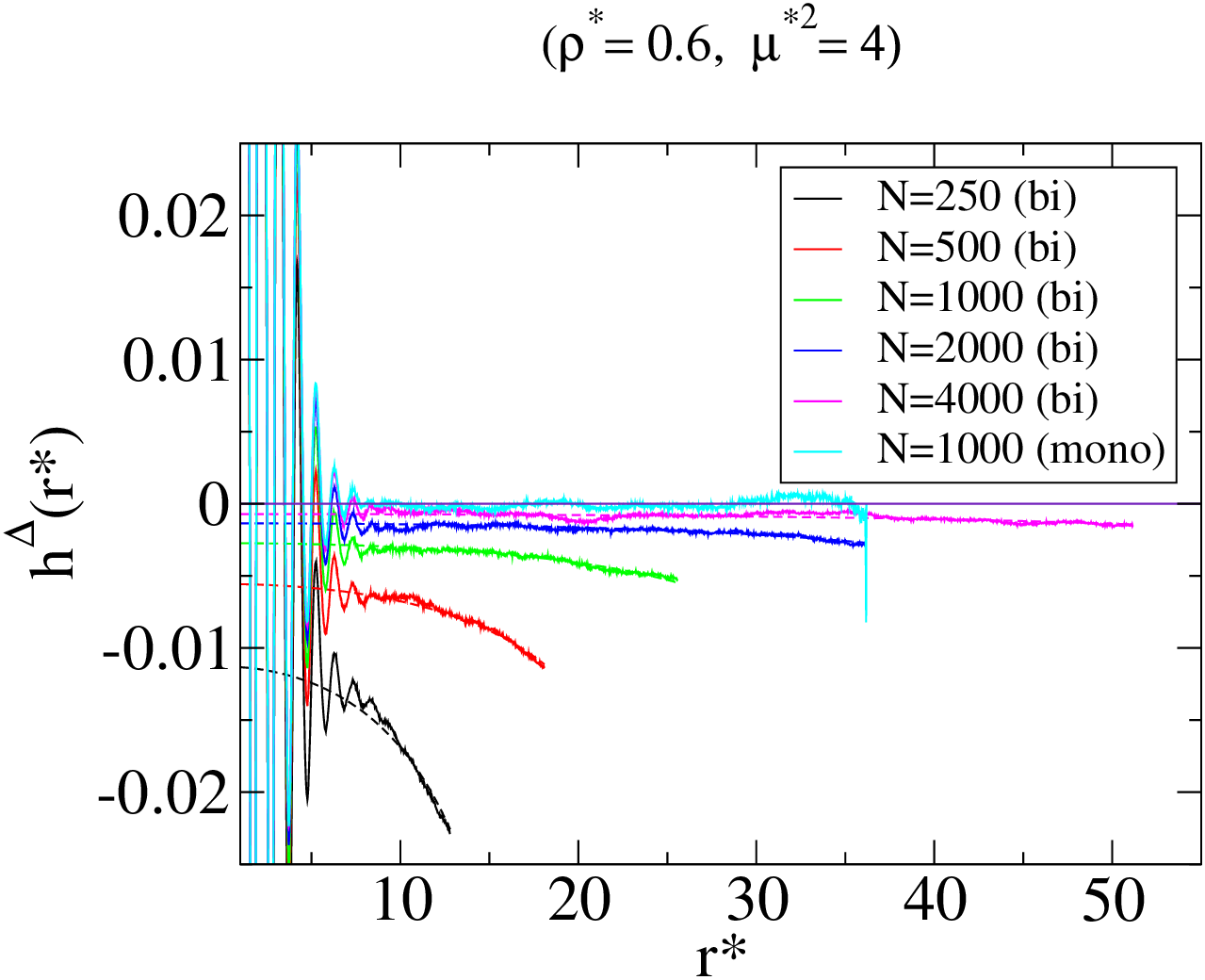}
\end{tabular}
\caption{Projection  $h^{\Delta}(r^*)$  for the states   I$\equiv$($\rho^{*} =0.7$, $\mu^{*2} = 2$) (top)
and   II$\equiv$($\rho^{*} =0.6$, $\mu^{*2} = 4$) (bottom)
and various numbers $N$ of particles (bi-dipoles) and $N=1000$ (mono-dipoles). 
Solid lines : MC data, dashed lines : asymptotic behavior.}
\label{Fig4}
\end{figure}

\begin{figure}[t!]
\centering
\begin{tabular}{ c c c }
\includegraphics[scale=0.45]{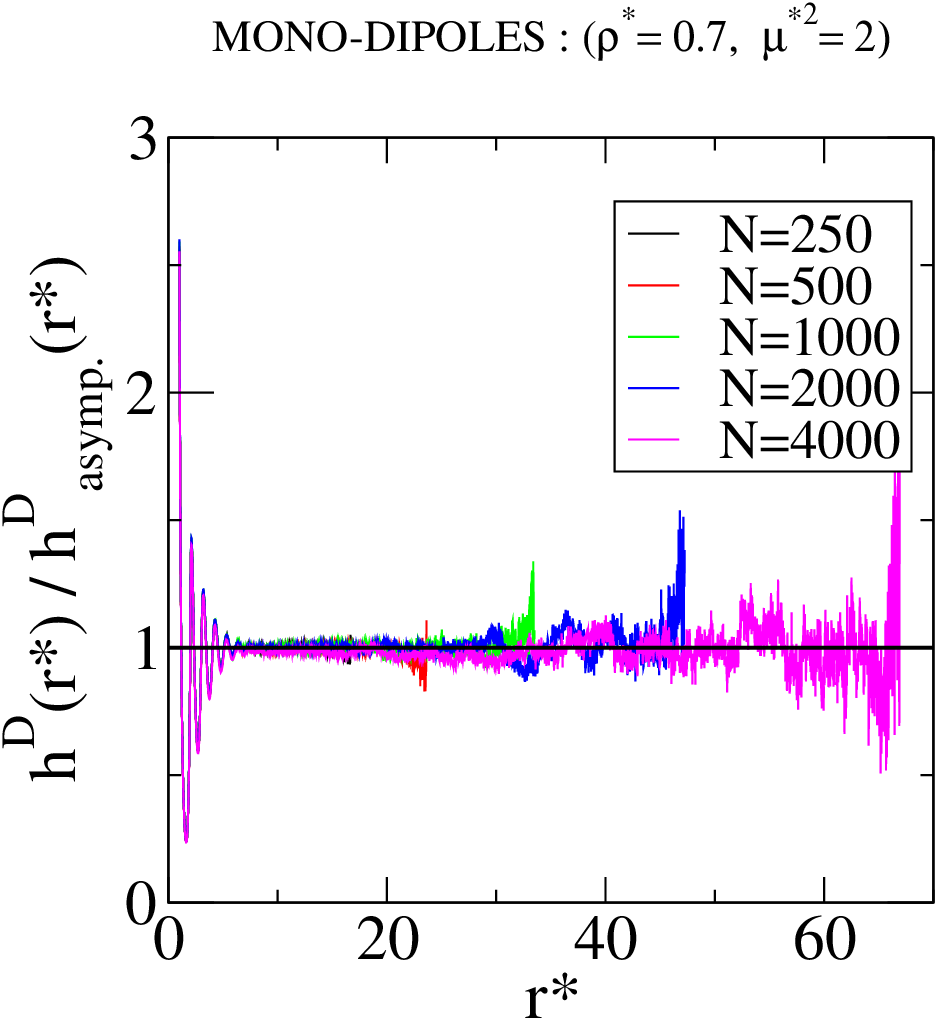} & \hspace{2em} & \includegraphics[scale=0.45]{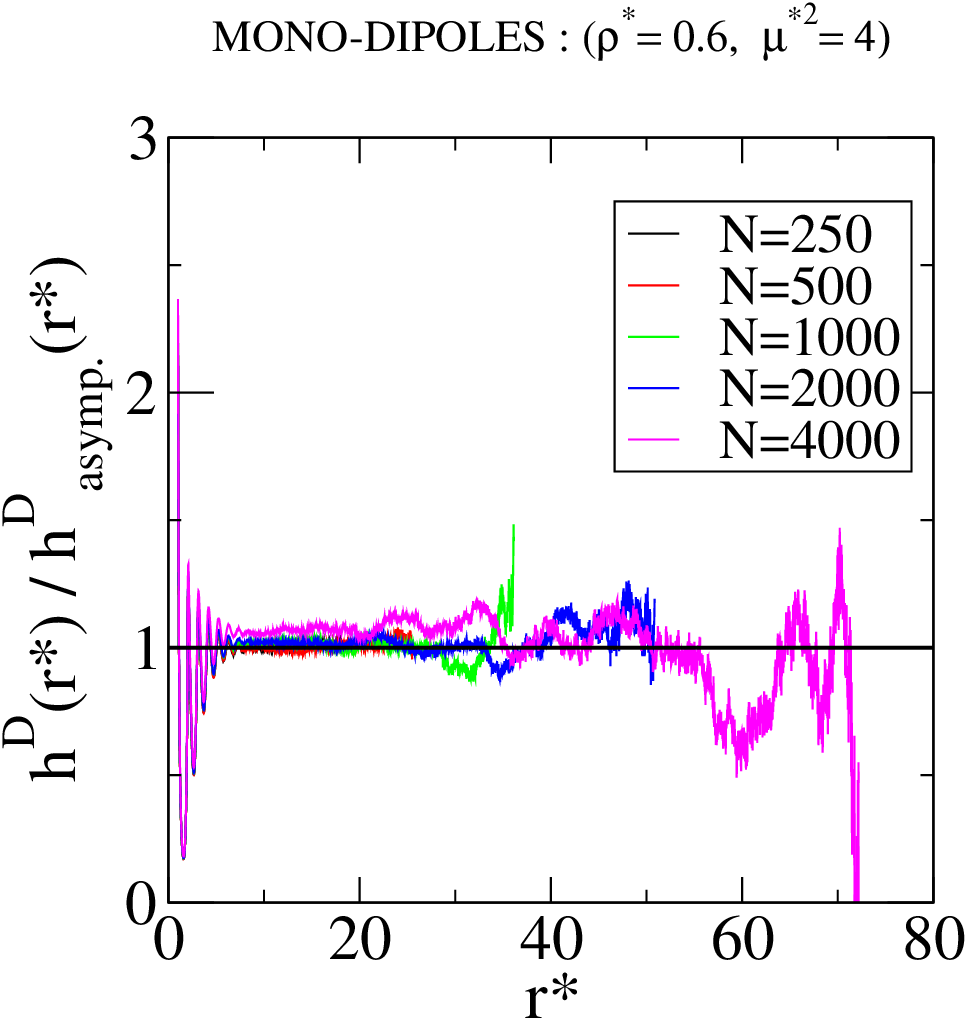} \\
\vspace{1.em} \\
\includegraphics[scale=0.45]{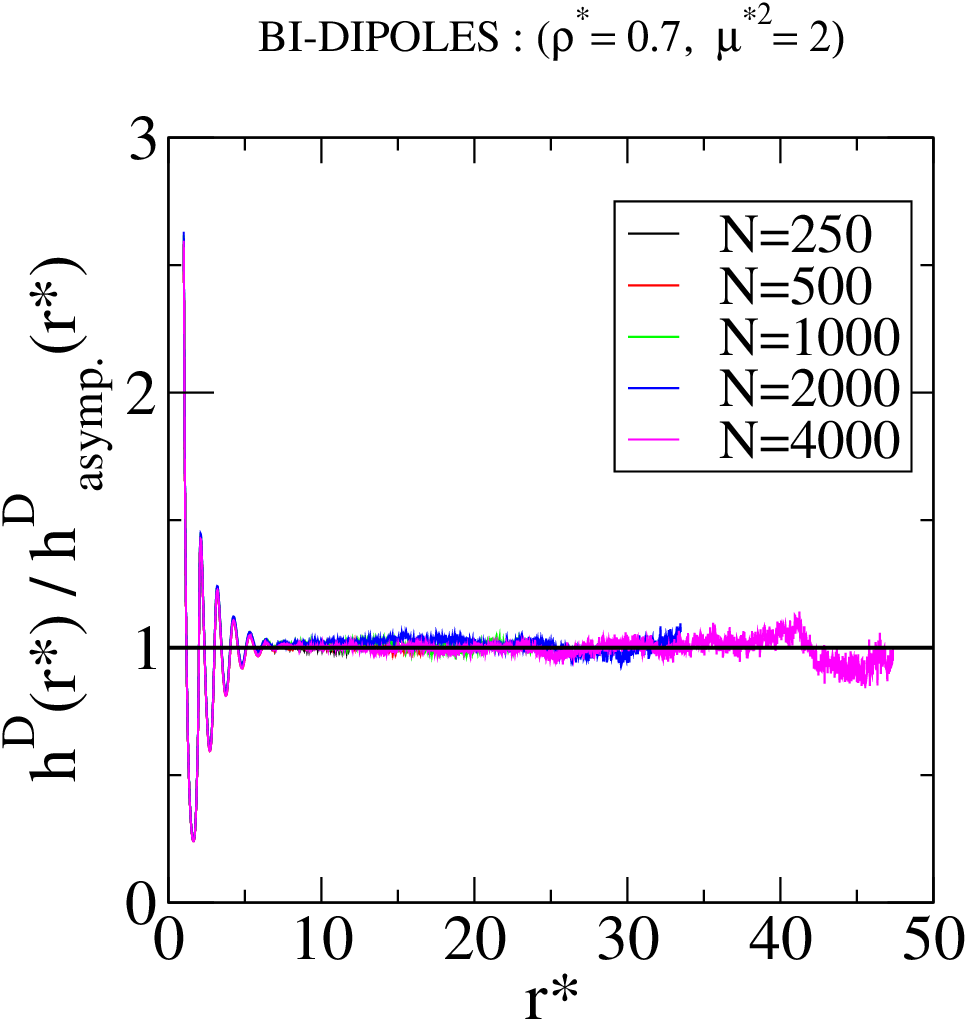}  & \hspace{2em} & \includegraphics[scale=0.45]{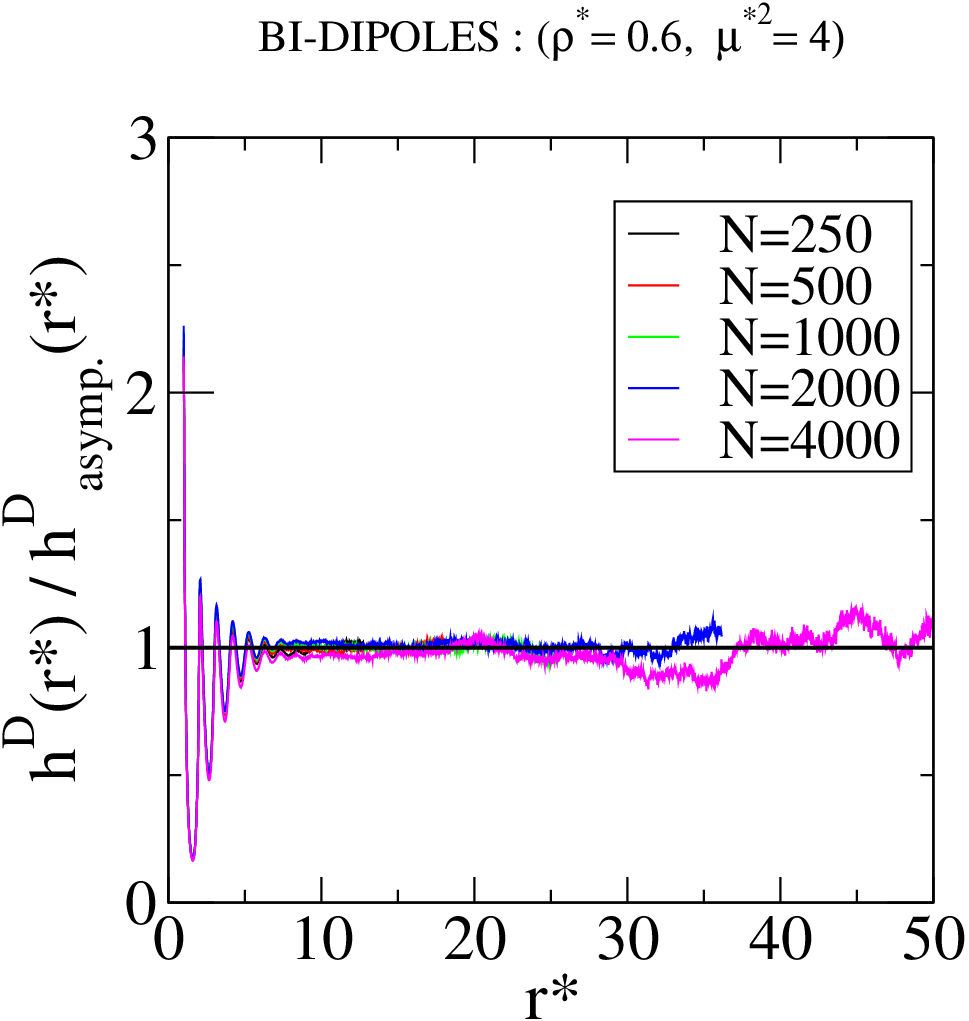}  
\end{tabular}
\caption{Ratio  $h^{D}(r^*)/h^{D}_{\mathrm{asymp.}}(r^*)$  for the state I$\equiv$($\rho^{*} =0.7$, $\mu^{*2} = 2$) (left)
and  II$\equiv$($\rho^{*} =0.6$, $\mu^{*2} = 4$) (right)
and various numbers $N$ of particles. 
Top : mono-dipoles, bottom : bi-dipoles.}
\label{Fig5}
\end{figure}

We check in  Fig.~\ref{Fig4} that the asymptotic behavior of the projection $h^{\Delta}(r)$
is indeed governed by the law of macroscopic electrostatics in $\mathcal{S}_2$, \textit{i.e}
by Eqs.~\eqref{assym_mono} for mono-dipoles and  Eqs.~\eqref{assym_bi} for bi-dipoles.
In all cases for  $r >\xi$ the asymptotic behavior of $h^{\Delta}(r)$ is in excellent agreement
with the theoretical prediction. Note that for mono-dipoles $h^{\Delta}(r)$ has no tail as expected from our theoretical
developments.
In the same vein Fig.~\ref{Fig5} displays the ratio $h^{D}(r)/h^{D}_{\mathrm{asymp.}}(r)$ for states~I and~II
and several system sizes. Once again a satisfactory agreement simulation/theory can be
observed. Note that for mono-dipoles (top curves) the noisy behavior at large distances
is a consequence of the divergence (due to an insufficient sampling)
of the term $\sim 1/\sin \psi_{ij}$ for $\psi_{ij} \sim \pi$
in the l.h.s. of Eq~\eqref{projo} which defines $h^{D}(r=R \psi)$ as a statistical average.

The fact that, for both states~I and~II,  the theoretical asymptotic behavior of $h^{\Delta}(r)$
and $h^{D}(r)$  is reproduced by the MC data is a clue that the system is in an isotropic liquid phase.
At lower temperatures  chains of dipoles, rings,  and topologically complex arrangements arise and the dielectric tensor
either does not exist or is no more local. This point will be discussed elsewhere.

\section{Conclusion}
\label{Conclu}
In this work we have clarified the laws of electrostatics on the sphere $\mathcal{S}_2(0,R)$ by 
introducing two types of basic elements : pseudo- and bi-charges from which two
types of multipoles can be obtained : mono- and bi-multipoles respectively.
The electrostatic potentials of these
multipoles can be obtained explicitely and exhaust all possible solutions of Laplace-Beltrami equation.

The application to polar fluids has been discussed in depth. Such a  fluid 
can  be represented either as an assembly of mono-dipoles living in $\mathcal{S}_2(0,R)$ 
or as a collection of  bi-dipoles living in the northern
hemisphere $\mathcal{S}_2(0,R)^+$. 

Since the dipolar Green's function are explicitely known for both
mono- and bi-dipoles,
Fulton's theory of dielectric fluids can be explicitely worked out in the two cases.
This includes the theory of the dielectric constant of an homogeneous and isotropic fluid and 
the derivation of the asymptotic tails of the projections 
$h^{\Delta}(r)$ and $h^{D}(r)$ of the equilibrium pair correlation function. It is likely that
a violation of these asymptotic laws signals a phase transition towards a
non-fluid phase.

We have reported and discussed MC  simulations of two states of the DHS fluid
which are both in the fluid phase and can serve as benchmark for future numerical works.
Finite size effects have been studied and allow to reach, at least for the internal energy, the thermodynamic
limit with a high precision ($\sim 10^{-5}$). Such a precision is smaller than the precision which could reasonably
be obtained on the dipolar Ewald potential and \textit{a fortiori} on  the mean internal energy  
which could be computed by an actual MC simulation which would make use of it. 
This suggests  that   $\mathcal{S}_2(0,R)$  is
a good geometry for precise MC simulations of the fluid phase. Simulations involving bi-dipoles appear
to attain more quickly the thermodynamic limit than simulations involving mono-dipoles.
An exploration of the phase diagram of the $2D$ DHS fluid will be presented elsewhere.

\begin{acknowledgments}
The author thanks  Jean-Jacques Weis and Martin Trulsson for interesting discussions.
\end{acknowledgments}

\appendix

\section{Mathematical properties of the dipolar Green's functions in $\mathcal{S}_2$}
\label{Appendix_A}

\subsection{Convolutions}
\label{convo}
First we consider the case of mono-dipoles. Let  $\mathbf{z}_1$ and $\mathbf{z}_2$ be two points of 
the sphere $\mathcal{S}_2$, we will show that
\begin{align}
\label{th1}
   \mathbf{G}_0^{\mathrm{mono}} \circ   \mathbf{G}_0 ^{\mathrm{mono}} (\mathbf{z}_1, \mathbf{z}_2)  &\equiv
 \ \int_{\mathcal{S}_2}d\, \Omega(\mathbf{z}_3) \; \mathbf{G}_0^{\mathrm{mono}}(\mathbf{z}_1, \mathbf{z}_3)
 \cdot \mathbf{G}_0^{\mathrm{mono}} (\mathbf{z}_3, \mathbf{z}_2)\nonumber \\
&= -\mathbf{G}_0^{\mathrm{mono}}(\mathbf{z}_1, \mathbf{z}_2)  \; ,
\end{align}
which shows that $-\mathbf{G}_0^{\mathrm{mono}}$ is a projector and has thus no inverse
with respect to the binary operator `` $\circ $  ''.
In order to prove Eq.~\eqref{th1} we rewrite the expansion~\eqref{G0_mono_b} of
the dipolar Green's function in spherical harmonics  as 
\begin{equation}
 \mathbf{G}_0^{\mathrm{mono}} (\mathbf{z}_1, \mathbf{z}_2) = \sum_{l=1}^{\infty}  \mathbf{G}_0^{l} (\mathbf{z}_1, \mathbf{z}_2)\; ,
\end{equation}
with
\begin{equation}
 \mathbf{G}_0^{l}(\mathbf{z}_1, \mathbf{z}_2)= -\frac{1}{l(l+1)} \sum_{m = -l}^{+l}
\nabla_{\mathcal{S}_2}\overline{Y}^{m}_{l}(\mathbf{z}_1) 
\nabla_{\mathcal{S}_2} Y_{l}^m (\mathbf{z}_2)  \; .
\end{equation}

To make further progress we need apply  Green-Beltrami theorem in $\mathcal{S}_2$
which states that~\cite{Atkinson} :
\begin{equation}
 \label{Green-Beltrami}
 \int_{\mathcal{S}_2} d\Omega \;  \nabla_{\mathcal{S}_2}f \cdot \nabla_{\mathcal{S}_2}g = - 
 \int_{\mathcal{S}_2} d\Omega \; f  \Delta_{\mathcal{S}_3}g \; ,
\end{equation}
where $f(\mathbf{z})$ and  $g(\mathbf{z})$ are functions defined on the unit sphere $\mathcal{S}_2$.
The  proof of theorem~\eqref{Green-Beltrami} is not so difficult and can be found, \textit{e.g.}
in the recent textbook by Atkinson and Han~\cite{Atkinson}.
It follows from Eq.~\eqref{Green-Beltrami} and the properties of spherical harmonics
that 
\begin{equation}
[\mathbf{G}_0^{l} \circ \mathbf{G}_0^{p}](\mathbf{z}_1, \mathbf{z}_2) = -\delta_{l,p} \,  \mathbf{G}_0^{l} (\mathbf{z}_1, \mathbf{z}_2)\;,
\end{equation}
from which the identity~\eqref{th1} is readily obtained.

The case of bi-dipoles is slightly different and has already been discussed in Ref.~\cite{Tr_Cai} for the $3D$  hyper-sphere 
 $\mathcal{S}_3$. We  now consider  two points $\mathbf{z}_1$ and $\mathbf{z}_2$ of the northern hemisphere
 $\mathcal{S}_2^+$. We will show that, again 
\begin{align}
\label{th2}
   \mathbf{G}_0^{\mathrm{bi}} \circ   \mathbf{G}_0 ^{\mathrm{bi}} (\mathbf{z}_1, \mathbf{z}_2)  &\equiv
 \ \int_{\mathcal{S}_2^+}d\, \Omega(\mathbf{z}_3) \; \mathbf{G}_0^{\mathrm{bi}}(\mathbf{z}_1, \mathbf{z}_3)
 \cdot \mathbf{G}_0^{\mathrm{bi}} (\mathbf{z}_3, \mathbf{z}_2)\nonumber \\
&= -\mathbf{G}_0^{\mathrm{bi}}(\mathbf{z}_1, \mathbf{z}_2)  \; .
\end{align}
Note  that, in this case,  the space integration is restricted to the northern hemisphere.
We follow the same strategy as for mono-dipoles and start with the expansion~\eqref{G0_bi_b} of
the dipolar Green's function  $\mathbf{G}_0^{\mathrm{bi}}(\mathbf{z}_1, \mathbf{z}_2)   $ in spherical harmonics
\begin{equation}
 \mathbf{G}_0^{\mathrm{bi}} (\mathbf{z}_1, \mathbf{z}_2) =2  \sum_{l \; odd}  \mathbf{G}_0^{l} (\mathbf{z}_1, \mathbf{z}_2)\; .
\end{equation}
For an odd value of $l$ we have $Y_{l}^m(-\mathbf{z}) = - Y_{l}^m(\mathbf{z})$ which implies that
\begin{equation}
\mathbf{G}_0^{l} \circ \mathbf{G}_0^{p}(\mathbf{z}_1, \mathbf{z}_2) = -\frac{1}{2} \, \delta_{l,p}  \, \mathbf{G}_0^{l} (\mathbf{z}_1, \mathbf{z}_2)\;,
\end{equation}
from which Eq.~\eqref{th2} follows.

In this Sec.~\ref{convo} we implicitely assumed that $R=1$. The reassessment of Eqs.~\eqref{th1} and~\eqref{th2} in
the case $R \neq 1$ is however trivial since the mono and bi dipolar Green's function 
$\mathbf{G}_0(1,2)$ scales as $R^{-2}$ with the radius of the sphere. Clearly Eqs.~\eqref{th1} and~\eqref{th2}
remain valid for  $R \neq 1$ with the replacement $d \Omega(\mathbf{z}) \to d S(M)$. 

\subsection{Integration over cones}
\label{cones}
This section is devoted  the integration of $\mathbf{G}_0^{\mathrm{mono}}
(\mathbf{z}_1, \mathbf{z}_2)$ on a cone of axis $\mathbf{z}_1$
and aperture $0 \leq \psi_0 \leq \pi$. We shall prove that
\begin{equation}
\label{int1}
 \int_{0 \leq \psi_{12}\leq \psi_0}d\, \Omega(\mathbf{z}_2) \; \mathbf{G}_0^{\mathrm{mono}}(\mathbf{z}_1, \mathbf{z}_2)=
   \dfrac{\cos \psi_0 -3 }{4}\, \mathbf{U}_{\mathcal{S}_2}(\mathbf{z}_1) \; .
\end{equation}
where $\psi_{12} =  \cos^{-1} ( \mathbf{z}_1 \cdot \mathbf{z}_2) $ is the angle between the two unit vectors 
$\mathbf{z}_1$ and $\mathbf{z}_2$.
To prove Eq.~\eqref{int1} one needs to take  some precaution because of the singularity of
$\mathbf{G}_0^{\mathrm{mono}}(\mathbf{z}_1, \mathbf{z}_2)$ at $\psi_{12} \to 0$. We make use
of the decomposition~\eqref{decompo} to rewrite
\begin{equation}
\label{int2}
 \int_{0 \leq \psi_{12}\leq \psi_0}d\, \Omega(\mathbf{z}_2) \; \mathbf{G}_0^{\mathrm{mono}}(\mathbf{z}_1, \mathbf{z}_2)
          = -\frac{1}{2}
 \mathbf{U}_{\mathcal{S}_2}(\mathbf{z}_1)
+ \lim_{\delta \to 0}  \int_{ \delta\leq \psi_{12}\leq \psi_0}d\, \Omega(\mathbf{z}_2) \; 
\mathbf{G}_0^{\mathrm{mono}}(\mathbf{z}_1, \mathbf{z}_2)\, .
\end{equation}
The integral $\mathbf{I}_{\delta}^{\psi_0}$ in the r.h.s. of Eq.~\eqref{int2} is computed by using spherical coordinates to reexpress the
formula~\eqref{G0_mono_b}
of the Green function and performing explicitely the integrals. 
One has 
$$
\mathbf{G}_0^{\mathrm{mono}}(\mathbf{z}_1, \mathbf{z}_2)=\frac{1}{4 \pi } \bigg\{
                                                                        \frac{1 + \cos \psi_{12}}{1 - \cos \psi_{12}}  
\mathbf{t}_{12} ( \mathbf{z}_1 ) \mathbf{t}_{12} (\mathbf{z}_2)
- \dfrac{1 }{1 - \cos \psi_{12}}
\mathbf{U}_{\mathcal{S}_2}(\mathbf{z_1}) \cdot  \mathbf{U}_{\mathcal{S}_2} (\mathbf{z_2}) \bigg\}
$$
By taking the north pole at point $\mathbf{z}_1$ we have 
the identifications  $\psi_{12} \to \theta$, $\mathbf{t}_{12}(\mathbf{z}_2) \to  \mathbf{e}_{\theta}(\theta, \varphi)$ and
 $\mathbf{t}_{12}(\mathbf{z}_1 )  \to  \mathbf{e}_{\theta}(\theta=0, \varphi) = (\cos \varphi, \sin \varphi, 0)^T$. Therefore
\begin{align}
\label{int3}
 \mathbf{I}_{\delta}^{\psi_0} &= \dfrac{1}{4 \pi} \int_{\delta}^{\psi_0} \sin \theta d \theta
\int_0^{2 \pi}  d \varphi  \bigg\{ 
 \frac{1 + \cos \theta}{1 - \cos \theta} \,  \mathbf{e}_{\theta}(\theta, \varphi) \mathbf{e}_{\theta}(0, \varphi) \nonumber \\
&- \dfrac{1 }{1 - \cos \theta} \mathbf{U}_{\mathcal{S}_2}(0, \varphi) \cdot  \mathbf{U}_{\mathcal{S}_2} (\theta, \varphi)
 \bigg\}  \nonumber \\
&=\dfrac{\cos \psi_0 - \cos \delta}{4}  \mathbf{U}_{\mathcal{S}_2} (\mathbf{z_1})  \; .
\end{align}
Taking the limit $\delta \to 0 $ and gathering the intermediate results~\eqref{int2} and~\eqref{int3} 
one  obtains the desired result~\eqref{int1}. 

The same type of calculation yields, for bi-dipoles, 
\begin{equation}
\label{int4}
 \int_{0 \leq \psi_{12}\leq \psi_0}d\, \Omega(\mathbf{z}_2) \; \mathbf{G}_0^{\mathrm{bi}}(\mathbf{z}_1, \mathbf{z}_2)=
( \frac{\cos \psi_{0}}{2} -1)
   \, \mathbf{U}_{\mathcal{S}_2}(\mathbf{z}_1) \; .
\end{equation}
with the restriction that   $0 \leq \psi_0 \leq \pi/2$.

\section{Fulton's theory in the plane.}
\label{Appendix_B}
In this section the domain occupied by the fluid consists of 
the entire $2D$ Euclidian plane $E_2$
with no boundaries at infinity.  We consider Fulton relation~\eqref{F_rel} in this geometry.
The susceptibility tensor $ \boldsymbol{\chi}(M_1, M_2)$   at thermal equilibrium is
\begin{subequations}
\begin{align}
 \boldsymbol{\chi}(M_1,M_2) &= 2 \pi \beta \mu^2\;
 \langle \sum_{i ,  j} \mathbf{s}_i   \mathbf{s}_j \,
 \delta^{(2)}\left( \mathbf{r}_1 - \mathbf{r}_i \right)
 \delta^{(2)}\left(\mathbf{r}_2 - \mathbf{r}_j\right)  \rangle  \; , \label{chi_a}\\
 &=\boldsymbol{\chi}_S(\mathbf{r}_{12}) 
                                    + \boldsymbol{\chi}_D(\mathbf{r}_{12})   \label{chi_b}\; ,
\end{align}
\end{subequations}
where $\overrightarrow{OM_i}=\mathbf{r}_i=(x_i,y_i)^T$ and  $ \mathbf{r}_{12} = \mathbf{r}_{2} - \mathbf{r}_{1}$.
The self-part $\boldsymbol{\chi}_S$ in~\eqref{chi_b} reads 
 \begin{equation}
  \boldsymbol{\chi}_S(\mathbf{r}_{12})  = y \mathbf{I}(\mathbf{r}_1, \mathbf{r}_2) \; ,
 \end{equation}
with $y=\pi \rho \beta \mu^2$  and 
$\mathbf{I}(\mathbf{r}_1, \mathbf{r}_2)=\mathbf{U} \delta^{(2)}(\mathbf{r}_{12})$
where  $\mathbf{U}= \mathbf{e}_x \mathbf{e}_x + \mathbf{e}_y\mathbf{e}_y$ is  the unit dyadic tensor  in $E_2 $.

The contribution $\boldsymbol{\chi}_D$ to Eq.~\eqref{chi_b} can be expressed
in terms of the pair correlation function 
$ \rho^{(2)}(1,2) \equiv \rho^{(2)}(\mathbf{r}_1,\alpha_1 ;\mathbf{r}_2,\alpha_2 )$ as
\begin{equation}
 \boldsymbol{\chi}_D (\mathbf{r}_{12}) = 2 \pi \beta \mu^2  \,
\int_0^{2 \pi}d\alpha_1 \;
\int_0^{2 \pi}d\alpha_2 \;
\rho^{(2)}(1,2) \; 
\mathbf{s}_1 \mathbf{s}_2 \; .
\end{equation}
For a homogeneous and isotropic fluid $  \rho^{(2)}(1,2) =(\rho/2 \pi)^2 g (1,2)$. The normalized
pair correlation function $g (1,2)$ can then be expanded
on the complete set of $2D$ orthogonal rotational invariants as in Refs~\cite{these,Orsay_1}
\begin{equation}
 h(1,2) = g(1,2) -1 = \sum_{m,n} h_{m,n}(r_{12}) \Phi_{mn} (1,2) \; ,
\end{equation}
where 
\begin{equation}
 \Phi_{mn} (1,2)=\cos(m \beta_1 + n \beta_2) \; .
\end{equation}
$\beta_1= \alpha_1 - \theta_{12}$ and $\beta_2= \alpha_2  - \theta_{12}$ are the angles of
the vectors $\mathbf{s}_1$ and  $\mathbf{s}_1$ and the direction $\mathbf{r}_{12}$ \cite{these,Orsay_1}
and the coefficients $h_{m,n}(r_{12})$ depends only on the sole modulus 
$r_{12}=\parallel \mathbf{r}_{12}\parallel$. The two invariants
$D(1,2)\equiv\Phi_{1 1} (1,2) $ (minus the angular part of the dipole-dipole energy
in the plane $E_2$)
and $\Delta(1,2)\equiv\Phi_{1 -1} (1,2) $ (scalar product $\mathbf{s}_1 \cdot \mathbf{s}_2$)
play an important role  since we have, as a short calculation will show
$$
 \boldsymbol{\chi}_D (\mathbf{r}_{12}) =\frac{\pi \beta \rho^2 \mu^2}{2} \left[
h^{\Delta}(_{12}) \mathbf{U}+
h^{D}(r_{12}) \left(2  \widehat{\mathbf{r}}_{12} \widehat{\mathbf{r}}_{12}    -\mathbf{U} \right)
\right] \; 
$$
where $ \widehat{\mathbf{r}}_{12}= \mathbf{r}_{12} /  r_ {12}$.
Gathering the intermediate results we obtain 
\begin{equation}
\label{chi_E2_final}
  \boldsymbol{\chi}(\mathbf{r}_{12}) = 
y \mathbf{I}(\mathbf{r}_{12})
+ \frac{y \rho}{2} h^{\Delta}(r_{12}) \mathbf{U}
+ \frac{y \rho}{2} h^{D}(r_{12})  \left(2  \widehat{\mathbf{r}}_{12} \widehat{\mathbf{r}}_{12}    -\mathbf{U} \right) \; .
\end{equation}

We turn now our attention to the dipolar Green's functions in $E_2$.  The bare Green's function is given by
 \begin{align}
\label{G0}
   \mathbf{G}_0(\mathbf{r}_1,\mathbf{r}_2) &=\frac{1}{2 \pi} \frac{\partial}{\partial \mathbf{r}_1}
\frac{\partial}{\partial \mathbf{r}_2} \log r_{12}  \; , \nonumber \\
                                                                &=\frac{1}{2 \pi}\frac{1}{ r_{12}^2}
\left(2  \widehat{\mathbf{r}}_{12} \widehat{\mathbf{r}}_{12}    -\mathbf{U} \right) \; . 
 \end{align}
It is important to remark that Eq.~\eqref{G0} implies that 
\begin{equation}
 \label{Trace_G0}
\Tr  \mathbf{G}_0(\mathbf{r}_1,\mathbf{r}_2) = \frac{1}{2 \pi} \Delta (-\log r_{12})
= -\delta^{(2)}(\mathbf{r}_{12}) \; .
\end{equation}

We note that in Fourier space $\widetilde{\mathbf{G}}_0 = - \widehat{\mathbf{k}}    \widehat{\mathbf{k}} $
(with $ \widehat{\mathbf{k}} =  \mathbf{k}/k $)  identifies with minus the projector
$\mathbf{P}_k \equiv  \widehat{\mathbf{k}}    \widehat{\mathbf{k}}$ on longitudinal modes. Denoting
by $\mathbf{Q}_k  = \mathbf{U} -   \mathbf{P}_k$ the projector on tranverse modes one has
the relations $\mathbf{P}_k \cdot \mathbf{P}_k = \mathbf{P}_k$, $\mathbf{Q}_k \cdot \mathbf{Q}_k = \mathbf{Q}_k$
and $\mathbf{P}_k \cdot \mathbf{Q}_k = 0$  from which
the identity
$(a \mathbf{P}_k + b\mathbf{Q}_k )^{-1} = a^{-1} \mathbf{P}_k + b^{-1}\mathbf{Q}_k  $ can be deduced.
Assuming  the locality of  dielectric tensor is local, \textit{i.e.}  $\widetilde{\boldsymbol{\epsilon}} = \epsilon \mathbf{U}$ in Fourier
space,  permits
 to compute the dressed  Green's function $\widetilde{\mathbf{G}}$. One gets, as expected, 
\begin{equation}
\label{Gtot}
 \widetilde{\mathbf{G}} = - \mathbf{P}_k \cdot \left[  \left(1 + \sigma \right) \mathbf{P}_k 
                                 + \sigma  \mathbf{Q}_k \right]^{-1} = \dfrac{\widetilde{\mathbf{G}}_0}{\epsilon}
\end{equation}

In order to obtain an expression for the dielectric constant we take the trace of both sides 
of Fulton's relation~\eqref{F_rel} and integrate over the whole plane. Making use of the expressions~\eqref{chi_E2_final}
for $\boldsymbol{\chi}$ and the formula~\eqref{Trace_G0} of the trace of  $\mathbf{G}_0$ one finds
\begin{equation}
 \label{epsilon}
\dfrac{(\epsilon -1)(\epsilon+1)}{2 \epsilon}=y \bigg(1+ \frac{\rho}{2}
\int_0^{\infty}2 \pi r h_{\Delta}(r) d r \; \bigg)
\end{equation}

Fulton's relation~\eqref{F_rel} gives asymptotically (\textit{i.e.} for $r_{12}>> \xi$, range of the dielectric tensor)
\begin{equation}
  \boldsymbol{\chi}(\mathbf{r}_{12})  \simeq \dfrac{(\epsilon -1)^2}{\epsilon}
\dfrac{1}{2 \pi r_{12}^2 } \left(2  \widehat{\mathbf{r}}_{12} \widehat{\mathbf{r}}_{12}    -\mathbf{U} \right) \; .
\end{equation}
A comparison of the above expression with Eq.~\eqref{chi_final} yields readily the asymptotic behaviors of 
$h_{D}(r)$ and $h_{\Delta}(r$ :
\begin{subequations}
\label{asy}
 \begin{align}
  h_{\Delta}^{\mathrm{asym.}}(r) &=0  \; , \\ 
  h_{D}^{\mathrm{asym.}}(r) &=\dfrac{(\epsilon-1)^2}{\epsilon}\dfrac{1}{\pi \rho y}\dfrac{1}{r^2} \; .
 \end{align}
\end{subequations}
from which we conclude that, while $ h_{\Delta}^{\mathrm{asym.}}(r)$ is a short range function,
the projection $h_{D}^{\mathrm{asym.}}(r)$ exhibits a long range algebraic asymptotic decay.
The results obtained in this appendix were obtained  by other methods
in Refs.~\cite{these,Orsay_1}. 
\section{Dipole displacements on the sphere  $\mathcal{S}_2$}
\label{Appendix_C}
The algorithm devised below adapts to the $2D$ case the algorithm used in Ref.~\cite{DIP3D}
for the $3D$ DHS fluid on the hypersphere $\mathcal{S}_3$.
The initial configuration of the Markov chain  is obtained  by sampling  $N$ vectors $\mathbf{z}^i$
uniformly on the sphere $\mathcal{S}_2$. The initial positions of the  point dipoles are thus
 $\mathbf{OM}^i = R\mathbf{z}^i $.  One uses the spherical  coordinates in the base $( \mathbf{e}_x ,
\mathbf{e}_y , \mathbf{e}_z ) $, \textit{i.e.}
$\mathbf{z}^i=( \sin \theta^i \cos \varphi^i, \sin \theta^i  \sin \varphi^i, \cos \theta^i) ^T  $ 
with $\cos \varphi^i = 2 \xi_1^i -1$ and $\varphi^i = 2 \pi \xi_2^i$ where
$\xi_1^i $ and $\xi_2^i $ are $2 N$ random numbers $\in (0,1)$. It is convenient
to complete the orthogonal local basis of spherical coordinates at point $M^i$  by defining also 
 $ \mathbf{u}^i = \partial \mathbf{z}^i /\partial\theta^i = (\cos \theta^i \cos \varphi^i, \cos \theta^i \sin \varphi^i, - \sin \theta^i)^{T}$ and
 $ \mathbf{v}^i =    \partial \mathbf{z}^i /\partial\varphi^i/ \sin \theta^i     =          (- \sin \varphi^i, \cos \varphi^i, 0)^{T}$.
The initial dipole $\boldsymbol{\mu}^i = \mu \mathbf{s}^i$ is  randomly sampled in the plane $\mathcal{T}(M^i)$
tangent to the sphere at point $M^i$ according to
$\mathbf{s}^i = \cos \phi^i \mathbf{u}^i + \sin \phi^i \mathbf{v}^i  $ with $\phi^i = 2 \pi \xi_3^i$ where
$\xi_3^i$ is  a random number $\in (0,1)$.

Due to the curvature of the space the trial move of dipole $\boldsymbol{\mu}^i$ is 
made in three steps  in which a displacement and a rotation are
involved. 
 First,  the new position $\mathbf{z}^i_{\mathrm{new}}$ of  the (randomly  chosen) dipole ``i''  is chosen uniformly
on a small cap of angle $\delta \theta$ about point $M^i$ (\text{i.e.} the intersection of a $3D$
cone of axis $\mathbf{z}^i$ and angle $\delta \theta$ with the sphere). It is convenient to use spherical
coordinates 
\textit{in the local basis}  $( \mathbf{z}^i,\mathbf{u}^i, \mathbf{v}^i ) $ so as
to define 
\begin{equation}
\mathbf{z}^i_{\mathrm{new}}= \sin  \theta^i_{\mathrm{new}} \cos \varphi^i_{\mathrm{new}}
\mathbf{u}^i + \sin \theta^i_{\mathrm{new}} 
  \sin \varphi^i_{\mathrm{new}}  \mathbf{v}^i     + \cos \theta^i_{\mathrm{new}}\mathbf{z}^i 
 \end{equation}
The choice $\cos\theta^i_{\mathrm{new}}= (1 -  \cos\delta \theta)\xi_4^i + \cos\delta \theta $ and $ \varphi^i_{\mathrm{new}}
= 2 \pi \xi_5^i$ where $\xi_4^i $ and $\xi_5^i $ are  random numbers $\in (0,1)$ ensures
a uniform sampling of the cap.
The new local basis at point $\mathbf{z}^i_{\mathrm{new}}$ is then given by
\begin{align}
 \mathbf{u}^i_{\mathrm{new}}& = \cos  \theta^i_{\mathrm{new}} \cos \varphi^i_{\mathrm{new}}
\mathbf{u}^i + \cos \theta^i_{\mathrm{new}} 
  \sin \varphi^i_{\mathrm{new}} \mathbf{v}^i   - \sin \theta^i_{\mathrm{new}}\mathbf{z}^i  \\
 \mathbf{v}^i_{\mathrm{new}}& = -\sin \varphi^i_{\mathrm{new}} \mathbf{u}^i 
 + \cos \varphi^i_{\mathrm{new}} \mathbf{v}^i \; . 
\end{align}

The second step consits in rotating the vector $\mathbf{s}^i$ in the tangent
plane  $\mathcal{T}(M_i)$ by an incremental angle $\delta \phi_i$ so that the new vector
 $\mathbf{s}^i_{\mathrm{Step 1}}$ reads as
\begin{equation}
\label{snew}
\mathbf{s}^i_{\mathrm{Step 1}} = \cos \delta \phi_i \mathbf{s}^i + \sin \delta \phi_i  \mathbf{z}^i \times \mathbf{s}^i \; ,
\end{equation}
where the choice $\delta \phi_i =  (\xi_6^i -0.5)  \delta \phi $ with $\xi_6^i $ some random number $\in (0,1)$
ensures a uniform sampling in the interval $(-\delta \phi /2,\delta \phi /2) $. 

However vector $ \mathbf{s}^i_{\mathrm{Step 1}}$ does not belong to the plane $\mathcal{T}(M^i_{\mathrm{new}})$ and the third and last step
of the trial MC move consists in a parallel transport of $ \mathbf{s}^i_{\mathrm{Step 1}} $
from the point $M^i$ to the new a priori position  $M^i_{\mathrm{new}}$ according
to Eq.~\eqref{transport} $ \mathbf{s}^i_{\mathrm{new}} = \tau_{M^iM^i_{\mathrm{new}}  } \mathbf{s}^i_{\mathrm{Step 1}}$,
\textit{i.e.}
\begin{equation}
  \mathbf{s}^i_{\mathrm{new}} = \mathbf{s}^i_{\mathrm{Step 1}} - 
\dfrac{  \mathbf{s}^i_{\mathrm{Step 1}} \cdot \mathbf{z}^i_{\mathrm{new}}}{1 + \cos \theta^i_{\mathrm{new}}}
(  \mathbf{z}^i + \mathbf{z}^i_{\mathrm{new}}) \; .
\end{equation}

The trial displacement of dipole $\mathbf{s}^i$ is then accepted or rejected according to the standard Metropolis 
criterion~\cite{Metropolis}
after testing the overlaps and the energies.
If the trial move is accepted we make the substitutions $\mathbf{z}^i_{\mathrm{new}} \rightarrow \mathbf{z}^i$,
 $\mathbf{u}^i_{\mathrm{new}} \rightarrow \mathbf{u}^i$,
$\mathbf{v}^i_{\mathrm{new}} \rightarrow \mathbf{v}^i$, and
$\mathbf{s}^i_{\mathrm{new}} \rightarrow \mathbf{s}^i$. 
In the numerical experiments reported in Sec.~\ref{Simulations} the parameters $\delta \theta $ and $\delta \phi$
were chosen in such a way that the acceptance ratio was $\sim 0.5$.

\newpage
\begin{table}[h!]
\centering
\begin{ruledtabular}
\begin{tabular}{ c c c c c c c }
$N$  & $\beta u^{\rm mono}$          & $\epsilon^{\rm mono}$ & $g^{00 \; {\rm mono}}(\sigma)$   & $h^{\Delta \; {\rm mono}}(\sigma)$   & $h^{D \; {\rm mono}}(\sigma)$         & $ Z^{\rm mono} $      \\
\hline 
250   &  -1.78533(8)     &    18.14(3)    &  4.335(1)    & 2.821(2) &  4.580(2) &   3.954(2)   \\
500   &  -1.78765(7)     &    18.09(3)    &   4.340(1)   & 2.819(2) &  4.582(2) &   3.970(2)  \\
1000 &  -1.78876(8)     &    17.98(5)    &   4.341(1)   & 2.823(2) &  4.577(2) &   3.977(2)  \\
2000 &  -1.78946(8)     &    17.93(7)    &   4.342(1)   & 2.817(2) &  4.575(2) &   3.981(2) \\
4000 &  -1.78969(8)     &    18.22(10)     &    4.339(1)  & 2.816(2) &  4.571(2) &   3.980(2) \\
$\infty$ & -1.78999(5)   &           -                     & - & -& -& -  \\
\hline
\hline
$N$  & $\beta u^{\rm bi}$          & $\epsilon^{\rm bi}$ & $g^{00 \; {\rm bi}}(\sigma)$   & $h^{\Delta \; {\rm bi}}(\sigma)$   & $h^{D \; {\rm bi}}(\sigma)$         & $ Z^{\rm bi} $      \\
\hline 
   250   &  -1.79010(7)      &  18.06(2)   &  4.334(1) & 2.800(2) & 4.583(2) &  3.962(2) \\
   500   &  -1.78998(8)      &  18.01(3)   &  4.339(1) & 2.811(2) & 4.579(2) &  3.974(2) \\
   1000 &  -1.79003(8)      &  17.83(5)  &  4.339(1) & 2.814(2) & 4.575(2) &   3.974(2)   \\
   2000 &  -1.79005(8)      &  17.75(7)   &  4.338(1)  & 2.814(2) & 4.569(2) & 3.978(2)  \\
   4000 &  -1.79000(8)      &  17.96(10)   &   4.334(1) & 2.809(2) & 4.562(2) & 3.975(2) \\
$\infty$ &  -1.79000(6)      &           -                    & -& -& -& - \\
\end{tabular}
\end{ruledtabular}
\caption{\label{I} Number of particles $N$, reduced internal energy per particle  $\beta u$, 
dielectric constant $\epsilon$, contact values $g^{00}(\sigma)$,  $h^{\Delta}(\sigma)$, and $h^{D}(\sigma)$
of some  projections 
of the  pair correlation function,
and compressibility factor $Z=\beta P / \rho$
of the DHS fluid
in the state \mbox{($\rho^{*}=0.7$, $\mu^{*2}=2$)} for
mono-dipoles (top) and   bi-dipoles (bottom). 
The number in bracket  corresponds to {\em two standard deviations}  and gives the accuracy of the last digits.
The thermodynamic limit of the internal energies were obtained 
from a linear regression in the variable  $N^{-1}$.
For each state
 $N_{\rm Conf.}=16.0 \times 10^9$ configurations were generated.}
\end{table}

\newpage
\begin{table}[h!]
\centering
\begin{ruledtabular}
\begin{tabular}{ c c c c c c c }
$N$  & $\beta u^{\rm mono}$          & $\epsilon^{\rm mono}$ & $g^{00 \; {\rm mono}}(\sigma)$   & $h^{\Delta \; {\rm mono}}(\sigma)$   & $h^{D \; {\rm mono}}(\sigma)$         & $ Z^{\rm mono} $      \\
\hline 
250     &   -4.1625(3)     &     43.94(9)       &  5.143(2)    & 5.531(4)   &   7.146(3)  &   1.660(2)  \\
500      &   -4.1670(3)    &     44.36(13)       &  5.142(2)    & 5.523(3)   &  7.135(3)   &   1.667(2)  \\
1000    &    -4.1693(3)   &    42.99(18)      &  5.136(2)    &  5.513(4)  &   7.123(3)  &   1.665(2)  \\
2000    &    -4.1703(3)   &     42.57(26)     &  5.133(2)    &  5.507(4)  & 7.117(3)    &   1.664(2)  \\
4000    &   -4.1710(3)    &     41.06(36)     &  5.123(2)    & 5.495(4)   & 7.099(3)    &   1.656(2)  \\
$\infty$ &  -4.17148(16) &           -             & - & -& -& -  \\
\hline
\hline
$N$  & $\beta u^{\rm bi}$          & $\epsilon^{\rm bi}$ & $g^{00 \; {\rm bi}}(\sigma)$   & $h^{\Delta \; {\rm bi}}(\sigma)$   & $h^{D \; {\rm bi}}(\sigma)$         & $ Z^{\rm bi} $      \\
\hline 
   250    &    -4.1722(3)      &     44.55(9)   &  5.136(2) & 5.501(3)  & 7.141(3)   &  1.657(2)   \\
   500    &     4.1718(3)      &     43.96(11)   &  5.136(2) & 5.506(4)  & 7.131(3)   &  1.657(2)   \\
   1000  &    -4.1717(3)      &    43.24(18)  &  5.132(2) & 5.502(4)  & 7.117(3)   &  1.662(2)   \\
   2000  &    -4.1714(3)      &    42.89(27)  &  5.122(2) & 5.487(4)  &  7.103(3)  &   1.655(2)  \\
   4000  &     -4.1714(3)     &    45.14()  &  5.137(2) & 5.510(4)  &  7.123(3)  &   1.670(2)  \\
$\infty$  &   -4.17138(15)   &           -                    & -& -& -& - \\
\end{tabular}
\end{ruledtabular}
\caption{\label{II}
Same as Table~\ref{I} but  for the state  \mbox{($\rho^{*}=0.6$, $\mu^{*2}=4$)}.}
\end{table}

\newpage
\newpage
\newpage
\newpage
\newpage

\end{document}